\documentclass[10pt,onecolumn]{IEEEtran}
\usepackage{url}
\usepackage[cmex10]{amsmath}
\usepackage{amsfonts}
\usepackage{amssymb}
\usepackage{pgf}
\usepackage{graphicx}
\usepackage{grffile}
\usepackage{subfig}
\usepackage{epstopdf}
\usepackage{mathtools}
\usepackage{mathabx} % for \not\perp

\ifCLASSINFOpdf
\else
\fi
\begin{document}
\title{The Hilbert spectrum and the Energy Preserving Empirical Mode Decomposition}
\author{Pushpendra Singh,
        Shiv Dutt Joshi,
        Rakesh Kumar Patney, and~Kaushik Saha % <-this % stops a space
\thanks{Pushpendra Singh, Shiv Dutt Joshi and Rakesh Kumar Patney are with the Department of Electrical Engineering, Indian Institute of Technology, Delhi,
Hauz Khas, New Delhi-110 016, INDIA e-mails: (pushpendra.singh@ee.iitd.ernet.in; sdjoshi@ee.iitd.ac.in; rkpatney@ee.iitd.ac.in).}% <-this % stops a space
\thanks{Kaushik Saha is with Samsung R \& D Institute India - Delhi, India e-mail: (kaushik.s14@samsung.com).}}% <-this % stops a space
%\thanks{Manuscript received April 19, 2005; revised December 27, 2012.}}
%\author{\IEEEauthorblockN{Pushpendra Singh\IEEEauthorrefmark{1}, Shiv Dutt Joshi\IEEEauthorrefmark{1}, Rakesh Kumar Patney\IEEEauthorrefmark{1} and Kaushik Saha\IEEEauthorrefmark{2}}
%\IEEEauthorblockA{\IEEEauthorrefmark{1}\\Department of Electrical Engineering, Indian Institute of Technology, Delhi\\
%Emails: pushpendra.singh@ee.iitd.ernet.in; rkpatney@ee.iitd.ac.in; sdjoshi@ee.iitd.ac.in}
%\IEEEauthorblockA{\IEEEauthorrefmark{2}\\Samsung R \& D Institute India - Delhi, India\\
%Email: kaushik.s14@samsung.com}}
%\markboth{Journal of \LaTeX\ Class Files,~Vol.~11, No.~4, December~2012}%
%{Shell \MakeLowercase{\textit{et al.}}: Bare Demo of IEEEtran.cls for Journals}
\markboth{Digital Signal Processing}%
{Singh P. \MakeLowercase{\textit{et al.}}: Energy Preserving Empirical Mode Decomposition}
\maketitle
\begin{abstract}
In this paper, we propose algorithms which preserve energy in empirical mode decomposition (EMD), generating finite $n$ number of band limited Intrinsic Mode Functions (IMFs). In the first energy preserving EMD (EPEMD) algorithm, a signal is decomposed into linearly independent (LI), non orthogonal yet energy preserving (LINOEP) IMFs and residue (EPIMFs). It is shown that a vector in an inner product space can be represented as a sum of LI and non orthogonal vectors in such a way that Parseval's type property is satisfied.
From the set of $n$ IMFs, through Gram-Schmidt orthogonalization method (GSOM), $n!$ set of orthogonal functions can be obtained. In the second algorithm, we show that if the orthogonalization process proceeds from lowest frequency IMF to highest frequency IMF, then the GSOM yields functions which preserve the properties of IMFs and the energy of a signal. With the Hilbert transform, these IMFs yield instantaneous frequencies and amplitudes as functions of time that reveal the imbedded structures of a signal. The instantaneous frequencies and square of amplitudes as functions of time produce a time-frequency-energy distribution, referred as the Hilbert spectrum, of a signal. Simulations have been carried out for the analysis of various time series and real life signals to show comparison among IMFs produced by EMD, EPEMD, ensemble EMD and multivariate EMD algorithms. Simulation results demonstrate the power of this proposed method.
\end{abstract}
\IEEEpeerreviewmaketitle
\section{Introduction}
%\hfill December 27, 2012
\IEEEPARstart{T}{he} empirical mode decomposition (EMD) is an adaptive signal analysis algorithm, introduced in \cite{rs1} for the analysis of non-stationary signals as well as signals generated from nonlinear systems and has become an established method for the signal analysis in various applications. The EMD decomposes a given signal into a finite number of band limited intrinsic mode functions (IMFs) which are derived directly from the data, unlike other signal decomposition techniques (like Fourier, Wavelets, etc.) which use predefined fixed basis for the signal analysis. The notion of instantaneous frequency (IF) and amplitude, derived from the Hilbert-Huang transform (HHT) provides an insight into the time-frequency and energy features of the signal. The Ensemble EMD (EEMD) is a noise-assisted data analysis method, developed in \cite{rs4}, to overcome the time scale separation problem of EMD. The Multivariate EMD (MEMD), developed in \cite{rs5}, is a generalization of the EMD for multichannel data analysis. The Compact EMD (CEMD) algorithm is proposed in \cite{rs12} to reduce mode mixing, end effect, and detrend uncertainty present in EMD and to reduce computation complexity of EEMD as well. To restrain the end effects and also to remove iterative errors and noise signal, wavelet analysis is used in the sifting process of the EMD, and a new stopping criterion based on correlation analysis is also proposed in \cite{rs25}. The IMFs generated by EMD are dependent on distribution of local extrema of signal and the type of spline used for upper and lower envelope interpolation and the traditional EMD uses cubic spline for upper and lower envelope interpolation. The EMD algorithm, proposed in \cite{rs13} to reduce mode mixing and detrend uncertainty, uses nonpolynomial cubic spline interpolation to obtain upper and lower envelopes, and have shown \cite{rs14} that it improves orthogonality among IMFs. Some recent studies, on the EMD based method have been performed for noise elimination~\cite{rs141} and condition-based adaptive trend prediction for rotating bearings~\cite{rs142}.

To eliminate energy leakage among IMFs, the Orthogonal EMD (OEMD) is proposed in \cite{rs3}, which generates orthogonal IMFs from the set of IMFs through the Gram-Schmidt orthogonalization method (GSOM). IMFs generated from EMD, EEMD and MEMD are not exactly orthogonal and hence there is always some energy leakage among the IMF components and the total sum of energies of IMFs is not equal to energy of signal, i.e. energy is not preserved in decomposition.

In any signal decomposition including EMD, the energy preserving property is important for the accurate and faithful, analysis and processing of three dimensional time-frequency distribution of the energy. To preserve the energy of a signal in the decomposition, we propose two EPEMD algorithms. First EPEMD algorithm directly provides the LI, non orthogonal yet Energy Preserving (LINOEP) IMFs and residue (EPIMFs), and we present this novel class of `LINOEP' functions in a well-posed mathematical result.
In the second EPEMD algorithm, to completely eliminate energy leakage among the IMF components, we propose a method to obtain `orthogonal' and `orthogonal \& uncorrelated' IMFs.

This paper is organized as follows: In section II we present brief review of the various variants of the EMD (i.e. EMD, EEMD, MEMD) and IMFs that are required in the present work. We propose the first EPEMD algorithm in section III. In section IV, the GSOM and an orthogonal EMD (OEMD) is discussed, and we propose the second EPEMD algorithm, through reverse order methodology, to obtain `orthogonal' and `orthogonal \& uncorrelated' IMFs. Simulation results are presented in section V. Section VI presents conclusions.
\section{The Empirical mode decomposition}
The EMD can decompose a stationary or non-stationary signal into a set of finite band-limited IMFs. The steps involved in EMD algorithm \cite{rs2}, to extract IMFs and residue from a given signal $x(t)$, are summarized in Algorithm 1.
The sifting process will be continued until the final residue is either a constant function, or a monotonic function, or a function with only one maximum and one minimum from which no more IMF can be derived. The decomposed signal $x(t)$ is expressed as the sum of $n$ IMF components plus the final residue:
\begin{equation}
 x(t)=\sum_{i=1}^{n}{y}_{i}(t) + r_n(t) =\sum_{i=1}^{n+1}{y}_{i}(t)\label{emd1}
\end{equation}
where $y_{i}(t)$  is the $i^{th}$ IMF and $r_n(t)=y_{n+1}(t)$ is final residue.
First IMF contains the finest scale or the shortest-period (i.e. highest frequency) oscillation and last IMF contains the longest-period (i.e. lowest frequency) oscillation present in the signal.
\\
\begin{tabular}{p{0.99\textwidth}}
\hline
Algorithm 1: Algorithm for EMD, for $i=1,\cdots,n$\\
\hline
${1.}$ Set $y_i(t)=x(t)$.
\\${2.}$ Obtain local maxima of the signal $y_i(t)$ and generate the upper envelope $e_u(t)$ by connecting the maxima with cubic spline interpolation.
\\${3.}$ Obtain local minima of the signal $y_i(t)$ and generate the lower envelope $e_l(t)$ by connecting the minima with cubic spline interpolation.
\\${4.}$ Obtain the mean signal $m(t)\triangleq[e_u(t)+e_l(t)]/2$.
\\${5.}$ Set $y_i(t) = y_i(t) - m(t)$ and determine if $y_i(t)$ is an IMF or not by checking the properties of IMF.
 Repeat step 2 to 5 and end when $y_i(t)$ is an IMF, and store it.
\\${6.}$ Set $x(t)=x(t)-y_i(t)$.
 \\${7.}$ Repeat step 1 to 6 and end when all the IMFs and residue of signal $x(t)$ are obtained.\\
\hline
\end{tabular}

\noindent The IMFs admit amplitude-frequency modulated (AM-FM) representation \cite{rs17} (i.e. $y_i(t) \approx a_i(t)\cos(\phi_i(t))$, with $a_i(t),\frac{d\,\phi_i(t)}{dt}>0$ $\forall t$) and well-behaved Hilbert transforms \cite{rs1}.
For any IMF $y_i(t)$, its Hilbert transform $\hat{y_i}(t)$ is defined as convolution of $y_i(t)$ and $1/\pi t$, i.e.
$\hat{y_i}(t)=\frac{1}{\pi}\int_{\infty}^{\infty}\frac{y_i(\tau)}{t-\tau} d\tau $ and the Hilbert transform emphasizes the local properties of $y_i(t)$.
An analytic signal $z_i(t)$ can be represented by
%\begin{equation}
 $z_i(t)=y_i(t)+j\hat{y_i}(t)=a_i(t)e^{j\phi_i(t)}$
%\end{equation}
where $a_i(t)=[y_i^2(t)+\hat{y_i}^2(t)]^{1/2}$, and  $\phi_i(t)=\tan^{-1}[\hat{y_i}(t)/y_i(t)]$ are instantaneous amplitude and phase of $y_i(t)$.
 The IF of $y_i(t)$ is defined as:
 %\begin{equation}
  $\omega_i(t)=\frac{d\,\phi_i(t)}{dt}=\frac{\frac{d\,\hat{y_i}(t)}{dt}y_i(t)-\hat{y_i}(t)\frac{d\,y_i(t)}{dt}}{\hat{y_i}^2(t)+y_i^2(t)}$.
  %\end{equation}
The physical meaning of IF $\omega_i(t)$ constrains that $\phi_i(t)$ must be a mono-component function of time, and the Bedrosian and Nuttall theorems \cite{rs20}, \cite{rs21} impose non-overlapping spectra constraints on the pair [$a_i(t),\cos(\phi_i(t))$].
An analytic representation of \eqref{emd1} is given by
 \begin{equation}
 z(t)=\sum_{i=1}^{n}{a}_{i}(t)\exp{(j\phi_i(t))}\label{emd2}
 \end{equation}
and for each IMF the IF $\omega_i(t)=\frac{d\,\phi_i(t)}{dt}$. For each IMF, the amplitude $a_i(t)$ and IF $\omega_i(t)$ are functions of time, and the three dimensional $\{t,\omega_i(t),a_i(t)\}$ time-frequency distribution of amplitude is Hilbert amplitude spectrum or Hilbert spectrum $H(\omega,t)$ or Hilbert-Huang spectrum (HHS). The marginal spectrum which is derived from Hilbert spectrum is defined as:
%\begin{equation}
$h(\omega)=\int_{0}^{T}H(\omega,t)dt$.
%\end{equation}
The marginal spectrum offers a measure of total amplitude (or energy) contribution from each value of frequency.
%\subsection{Intrinsic Mode Functions and their properties}

The IMF components obtained from EMD methods should follow the requirements of completeness, orthogonality, locality and adaptiveness.
The IMFs obtained from EMD satisfy the requirements of completeness, which means that the sum of the IMFs and residue reconstruct the original signal, and approximately follow the requirement of orthogonality \cite{rs1}.
All IMFs must satisfy two basic conditions: (1) In the complete range of time series, the number of extrema (i.e. maxima and minima) and the number of zero crossings are equal or differ at most by one. (2) At any point of time, in the complete range of time series, the average of the values of upper and lower envelopes, obtained by the interpolation of local maxima and the local minima, is zero.
The first condition ensure that IMFs are narrow band signals and the second condition is necessary to ensure that the IF does not have redundant fluctuations because of asymmetric waveforms.
%The IMFs are approximately orthogonal but sometimes there is severe energy leakage among IMFs.
The energy of any signal $x(t)$, defined over the time $[0,T]$, is given by $E_{x}=\int_{0}^T x^2(t) dt$
 and energy leakage between two IMFs can be calculated by (with $j,k=1,2,...,n, j\neq k$)
$E_{jk}=\int_{0}^T y_{j}(t)y_{k}(t) dt$.
An overall index of orthogonality, denoted as $IO_T$, and a partial index of orthogonality for any two IMFs components, denoted as $IO_{jk}$, are defined \cite{rs1} as follows:
\begin{equation}
IO_{T}\triangleq\frac{\displaystyle{\sum_{j=1}^{n+1}\sum_{\substack{k=1 \\ k \neq j}}^{n+1}}\int_{0}^T y_{j}(t)y_{k}(t) dt}{\int_{0}^T x^2(t) dt} \label{iot}
\end{equation}
\begin{equation}
IO_{jk}\triangleq\frac{\int_{0}^T y_{j}(t)y_{k}(t) dt}{\int_{0}^T y^2_{j}(t) dt +\int_{0}^T y^2_{k}(t) dt}
\end{equation}
The ideal values of energy leakage, overall as well as partial index of orthogonality are zero. The authors in \cite{rs1} observed that there is almost orthogonality among IMFs. The numerical simulations in \cite{rs3} demonstrated that the minor error in orthogonality, considered in \cite{rs1}, is not always valid, and there is actually severe energy leakage when EMD is applied for the decomposition of time series. The IMFs are not theoretically orthogonal, and hence the value of $IO_T$ is small but not zero, and sometimes very severe as shown in simulation results. The percentage error in energy ($Pee$) is defined as:
\begin{equation}
Pee\triangleq(\frac{E_x-E_{emd}}{E_x})\times100 \label{pee}
\end{equation}
where $E_{emd}\triangleq\sum_{i=1}^{n+1}\int_{0}^Ty^2_i(t)dt$ is sum of energies of IMFs and residue.
From \eqref{emd1}, \eqref{iot} and \eqref{pee} we obtain that:
\begin{equation}
Pee=IO_T\times100
\end{equation}
%\subsection{Ensemble Empirical Mode Decomposition}

The EMD may suffer from mode mixing, aliasing and end effect artefacts \cite{rs22}. To overcome these issues of EMD, a noise-assisted data analysis method EEMD, which derives the true IMF components as the ensemble average of all trials, from the signal added with a different realization of white noise of finite amplitude in each trials. %The EEMD utilizes the scale separation capability of the EMD to reduce the mode mixing problem, and is considered as a significant improvement over the EMD method.
The EEMD algorithm for the signal $x(t)$, to obtain IMFs, can be summarized as follows:
\\
\begin{tabular}{p{0.99\textwidth}}
\hline
Algorithm 2: Algorithm for EEMD\\
\hline
${1.}$ Add a realization of white noise $n_i(t) \sim \mathcal{N}(0,\sigma^2)$ to the signal $x(t)$ i.e. set $x_i(t) = x(t)+n_i(t)$.
\newline ${2.}$ Using EMD algorithm, obtain IMFs of the white noise added signal $x_i(t)$ and store them. Repeat step $1$ and step $2$ for $i=1,2,\cdots,N$ with different realization of white noise each time.
\newline ${3.}$ Obtain the ensemble average of corresponding IMFs and residue of the decompositions as the final result.\\
\hline
\end{tabular}
\\Due to ensemble averaging of corresponding IMFs and residue, added white noise cancel each other in the final mean, and the mean IMFs stay within the natural dyadic filter windows. Therefore, EEMD preserves the dyadic property of the IMFs of any data, and greatly reduces the probability of mode mixing.
%\subsection{Multivariate Empirical Mode Decomposition}

The EMD and EEMD are well suited for univariate signal and when applied channel-wise to multichannel signal analysis may suffer form nonuniformity, scale alignment and nature of IMFs \cite{rs22}. The Multivariate EMD (MEMD) is a generalization of the bivariate \cite{rs6} and trivariate \cite{rs7} EMD and overcomes these issues. The MEMD  generates multidimensional envelopes by mapping multivariate signal into multiple real-valued projected signals and yield multi-dimensional rotational modes via the corresponding multivariate IMFs.
Let, the column vectors of the matrix $\mathbf{X}=
[\begin{array}{cccc}
\mathbf{x}_1 &\mathbf{x}_2 &\cdots &\mathbf{x}_n
\end{array}]$ of dimension $p\times n$, represent an $n$-dimensional multivariate signal, and the row vector
$\mathbf{d}^{\theta^k}=
[\begin{array}{cccc}
d^{\theta^k}_1 &d^{\theta^k}_2  &\cdots &d^{\theta^k}_n
\end{array}]$
 represent a $k^{th}$ direction vector corresponding to the directions given by $k^{th}$ angle
 $\mathbf{\theta}^k=
[\begin{array}{cccc}
\theta^k_1 &\theta^k_2  &\cdots &\theta^k_{(n-1)}
\end{array}]$ on an $(n-1)$ sphere, for all $k=1$ to $K$ directions, respectively. The MEMD algorithm, which is extension of EMD for multichannel data analysis, suitable for decomposition of general nonlinear and non-stationary multivariate time series is summarized from \cite{rs5} as follows.
\\
\begin{tabular}{p{0.99\textwidth}}
\hline
Algorithm 3: Algorithm for MEMD\\
\hline
${1}.$ Set $\mathbf{M = X}$.
\\${2}.$ Obtain angle $\mathbf{\theta}^k$ corresponding to the linearly normalized low-discrepancy Hammersley sequences \cite{rs18} on $(n-1)$ sphere.
\\${3}.$ Obtain coordinates of unit direction vector $\mathbf{d}^{\theta^k}$ corresponding to angle calculated in step one.
\\${4}.$ Take projection of input signal on $k^{th}$ direction vector i.e. $\mathbf{p}^{\theta^k}=\mathbf{d}^{\theta^k}\bullet \mathbf{M}^T$.
\\${5}.$ Obtain the time instants of maxima and minima of the projected signal $\mathbf{p}^{\theta^k}$.
\\${6}.$ Generate multidimensional upper $\mathbf{U}_k$ and lower $\mathbf{L}_k$ envelopes, by spline interpolation of the input signal $\mathbf{X}$ at the time instants of maxima and minima, respectively.
\\${7}.$ Obtain mean of the multidimensional upper and lower envelopes i.e. $\mathbf{E}_k=[\mathbf{U}_k+\mathbf{L}_k]/2$ and set $\mathbf{E}_k=\mathbf{E}_{(k-1)}+\mathbf{E}_k$, where $\mathbf{E}_0$ is null matrix. Repeat the steps 2 to 7 for $k=1$ to $K$ directions and at the end set $\mathbf{E}_k=\mathbf{E}_k/K$.
\\${8}.$ Set $\mathbf{M = M - E}_k$ and determine if $\mathbf{M}$ is a multivariate IMFs or not.
 Repeat step 2 to 8 and end when $\mathbf{M}$ is a multivariate IMF, and store it.
\\${9}.$ Set $\mathbf{X= X-M}$. Repeat step 1 to 9 and end when all the multivariate IMFs and residue of signal $\mathbf{X}$ are obtained.\\
\hline
\end{tabular}
\\The stoppage criterion for multivariate IMFs is similar to that proposed in \cite{rs8}, and the first condition of IMF is not imposed as extrema cannot be properly defined for multivariate signals \cite{rs9}. The MEMD algorithm separates out common oscillatory modes presents within multivariate data which makes it suitable for stationary and non-stationary multichannel data analysis.

%\noindent\begin{minipage}{.5\textwidth}
\begin{figure}[!t]
\centering
\includegraphics[width=0.99\textwidth]{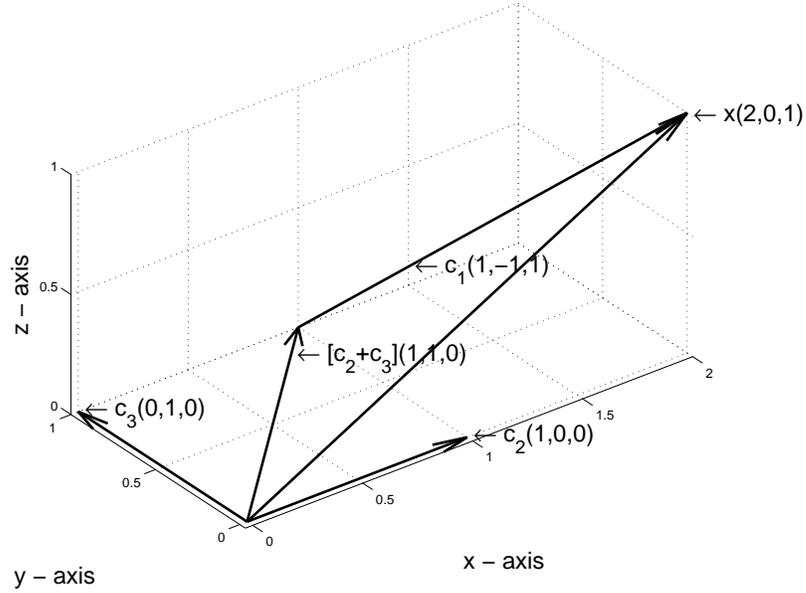}
\captionof{figure}{Three non orthogonal vectors $\mathbf{c_1,c_2,c_3}$ such that $\mathbf{c_1\perp(c_2+c_3)}$, $\mathbf{c_2\perp c_3}$ and vector $\mathbf{x=c_1+c_2+c_3}$ in 3-D.}
\label{fig:threedc123}
\end{figure}
%\end{minipage}
%\begin{minipage}{.5\textwidth}
\begin{figure}[!t]
\centering
\includegraphics[angle=0]{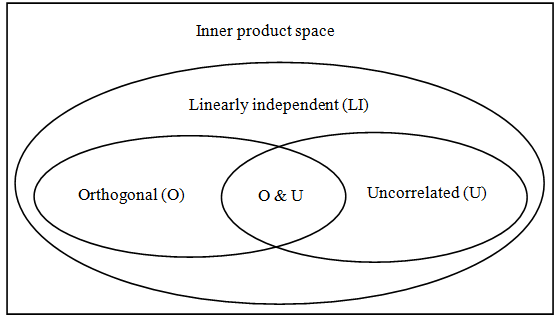}
\captionof{figure}{The inner product space}
\label{fig:Vspace}
\end{figure}
\section{The EPEMD and LINOEP Intrinsic Mode Functions}
The energy preserving property is important for a variety of reasons, and it is obtained by the orthogonal decomposition of signal in various transforms like Fourier, Wavelet, Fourier-Bessel, etc. The energy preserving property is especially important for the accurate and faithful analysis and processing of three dimensional time-frequency distribution of the energy of a signal. The EMD algorithm, inherently, neither ensures orthogonality nor does it preserve the energy of a signal in the decomposition.

The IMFs, generated through EMD, are not exactly orthogonal and hence energy is not preserved (i.e. signal energy is not equal to sum of individual component energy). The overall index of orthogonality is not zero, and hence there is, sometimes, very large percentage error in the energy as shown in simulation results. To completely preserve the energy of decomposition and to achieve zero percentage error in energy (i.e. to obtain ideal value `zero' of overall index of orthogonality), we propose first EPEMD algorithm to generate linearly independent (LI), non orthogonal yet energy preserving (LINOEP) IMFs and residue as follows.

Applying EMD to time series ${x(t)}$, one can write ${x(t) = y_1(t) + r_1(t)}$, where, nonzero and LI signals ${y_1}(t)$ and ${r_1}(t)$ are the first IMF and residue, respectively.
\\Let ${s_{11}(t)=r_1(t)}$ and ${s_{12}(t)=y_1(t)-\alpha_1 s_{11}(t)}$,
where ${\alpha_1=\frac{\langle y_1(t),s_{11}(t) \rangle }{\langle s_{11}(t),s_{11}(t) \rangle }}$ is such that ${s_{11}(t)}$ and ${s_{12}(t)}$ are orthogonal. Through addition of ${s_{11}(t)}$ and ${s_{12}(t)}$, we obtain
%\begin{equation}
$y_1(t) + r_1(t)=s_{12}(t)+(1+\alpha_1)s_{11}(t)$ $=c_1(t)+c'_2(t)$ i.e.
%\end{equation}
\begin{equation}
{x(t) = y_1(t) + r_1(t)=c_1(t)+c'_2(t)} \label{st12}
\end{equation}
where IMF ${c_1(t)=s_{12}(t)}$ is orthogonal to residue ${c'_2(t)=(1+\alpha_1)s_{11}(t)}$.
\\At the each stage of decomposition there are two permutations of IMF and residue, and hence two choices are available to perform the GSOM. We proceed from the residue (lower mode of oscillations) and orthogonalize IMF and residue to obtain proper IMF $c_1(t)$ and residue $c_2'(t)$ which preserve the properties of the IMF. If we proceed from IMF (higher mode of oscillations) to obtain orthogonal IMF and residue, then higher mode of oscillations get mixed up in residue.
\\Through EMD and the above procedure, ${c'_2(t)}$ and subsequent residues are further decomposed into two orthogonal components i.e.

\begin{equation}
{c'_2(t)=y_2(t) + r_2(t)=c_2(t)+c'_3(t)}
\ \text{ ; }{c_2(t) \perp c'_3(t)} \label{st22}
\end{equation}
\begin{equation}
{c'_3(t)=y_3(t) + r_3(t)=c_3(t)+c'_4(t)}
\ \text{ ; }{c_3(t) \perp c'_4(t)} \label{st32}
\end{equation}
%\begin{equation*}
%\vdots
%\end{equation*}
$\qquad \qquad \qquad \qquad \qquad \qquad \qquad \qquad \qquad \qquad \qquad \qquad \vdots$
\begin{equation}
{c'_{n}(t)=y_{n}(t)+r_{n}(t)=c_{n}(t)+c_{n+1}(t)}
\ \text{ ; }{c_{n}(t) \perp c_{n+1}(t)} \label{st42}
\end{equation}
where the residue $r_n(t)$ is not zero, if it is zero then we do not orthogonalize $y_n(t)$ and $r_n(t)$.
By using equations \eqref{st12} to \eqref{st42}, we can write ${x(t)}$ as:
\begin{equation}
{x(t) =\sum_{i=1}^{n}y_i(t)+r_n(t)=\sum_{i=1}^{n+1}c_i(t)} \label{st52}
\end{equation}
%At any decomposition level, the residue $r_i(t)$ is not zero, if it is zero then EMD algorithm stops at that level and we do not %orthogonalize $y_i(t)$ and $r_i(t)$.
In this decomposition, in general, ${c_i(t)} \not\perp {c_j(t)}$ for ${i,j=1,2,\dots,n}$, but we always have the condition that ${c_i(t) \perp \sum_{j=i+1}^{n+1}c_j(t)}$. Such an example of 3D vectors is shown in Figure~\ref{fig:threedc123}.
It is to be noted that this procedure can be easily extended for MEMD, to obtain energy preserving MEMD (EPMEMD), and we can derive energy preserving multivariate IMFs. Based on the above discussions on EPEMD and LINOEP IMFs, we propose following the mathematical result.
%\begin{theorem}\label{EPVThrm1}
\newtheorem{name}{Theorem}
\begin{name}
Let $H$ be a Hilbert space over the field of complex numbers, and let $\{\mathbf{x},\mathbf{x}_1,\cdots,\mathbf{x}_{n+1}\}$ be a set of vectors satisfying the following conditions:
%\\$(i)$ $\{\mathbf{x}_1,\cdots,\mathbf{x}_{n+1}\}$ are lineally independent (LI).
\begin{equation}
(i) \qquad \qquad \qquad \qquad  {\mathbf{x}_i \perp \sum_{j=i+1}^{n+1}\mathbf{x}_j} \label{st1}
\end{equation}
\begin{equation}
(ii) \qquad \qquad \qquad \qquad \qquad  \mathbf{x} =\sum_{i=1}^{n+1}\mathbf{x}_i \label{st2}
\end{equation}
Then in the representation, given in \eqref{st2}, the square of the norm, and hence energy is preserved, i.e.
\begin{equation}
\left\lVert \mathbf{x} \right\rVert^2= \left\lVert \sum_{i=1}^{n+1}\mathbf{x}_i \right\rVert^2 =\sum_{i=1}^{n+1}\lVert \mathbf{x}_i\rVert^2
\end{equation}
\end{name}
%\end{theorem}
\begin{IEEEproof}
 We prove the result using mathematical induction.
\\Base Case: For $n=1$, theorem is true, since $ \mathbf{x} = \mathbf{x}_1+\mathbf{x}_2$, with $\mathbf{x}_1 \perp \mathbf{x}_2$, which implies inner product $\langle \mathbf{x}_1,\mathbf{x}_2 \rangle =\langle \mathbf{x}_2,\mathbf{x}_1 \rangle =0$, and $\lVert \mathbf{x}\rVert ^2 =\langle \mathbf{x}_1+\mathbf{x}_2,\mathbf{x}_1+\mathbf{x}_2 \rangle$
\\$ = \langle \mathbf{x}_1,\mathbf{x}_1 \rangle +\langle \mathbf{x}_1,\mathbf{x}_2 \rangle +\langle \mathbf{x}_2,\mathbf{x}_1 \rangle +\langle \mathbf{x}_2,\mathbf{x}_2 \rangle$
\\$ =\lVert \mathbf{x}_1\rVert ^2+\lVert \mathbf{x}_2\rVert ^2$.
\\Induction hypothesis: Suppose that result is true for some $n = k\ge1$, that is, if $\mathbf{x}_i \perp \sum_{j=i+1}^{k+1}\mathbf{x}_j$ and $\mathbf{x} = \sum_{i=1}^{k+1}\mathbf{x}_i$, then
$\left\lVert \mathbf{x} \right\rVert^2= \left\lVert \sum_{i=1}^{k+1}\mathbf{x}_i \right\rVert^2 =\sum_{i=1}^{k+1}\lVert \mathbf{x}_i\rVert^2$.
\\Induction step: We show that the result is also true for $n = k + 1$, that is, if $\mathbf{x}_i \perp \sum_{j=i+1}^{k+2}\mathbf{x}_j$ and $\mathbf{x} = \sum_{i=1}^{k+2}\mathbf{x}_i$ then we have to show that $\left\lVert \mathbf{x} \right\rVert^2= \sum_{i=1}^{k+2} \left\lVert \mathbf{x}_i \right\rVert^2$.
\\We have $\mathbf{x}=\mathbf{x}_1+\mathbf{y}$, where $\mathbf{y}= \sum_{i=2}^{k+2}\mathbf{x}_i$. Since, $\mathbf{x}_1 \perp \mathbf{y}$, therefore, $\lVert \mathbf{x}\rVert ^2 =\lVert \mathbf{x}_1\rVert ^2+\lVert \mathbf{y}\rVert ^2$. From induction hypothesis we obtain $\left\lVert \mathbf{y} \right\rVert^2= \left\lVert \sum_{i=2}^{k+2}\mathbf{x}_i \right\rVert^2 =\sum_{i=2}^{k+2}\lVert \mathbf{x}_i\rVert^2$ (it's only a matter of indexing), and hence
%Recursively evaluating inner product by using ${\mathbf{x}_i \perp \sum_{j=i+1}^{k+2}\mathbf{x}_j}$, we obtain
%\\$\lVert \mathbf{x}\rVert ^2 =\lVert \mathbf{x}_1\rVert ^2+\lVert \mathbf{x}_2\rVert ^2+\cdots+ \lVert \mathbf{x}_{k+2}\rVert ^2$
\\$\lVert \mathbf{x}\rVert ^2 =\sum_{i=1}^{k+1}\left\lVert \mathbf{x}_i \right\rVert^2+ \lVert \mathbf{x}_{k+2}\rVert ^2$. Hence by
mathematical induction theorem is true for all positive integers $n$, which completes the proof.
\end{IEEEproof}
%This result is valid for finite-dimensional ($\mathbb{R}^m$, $\mathbb{C}^m$) as well as infinite-dimensional ($\mathbb{L}^2$) Hilbert spaces.
Discussion: The following observations, regarding the above result, are made:
\\(1) This is a signal specific decomposition, i.e., $\mathbf{x} =\sum_{i=1}^{n+1}\mathbf{x}_i$. \\(2) This decomposition is not unique and there may be so many sets of $\{\mathbf{x},\mathbf{x}_1,\cdots,\mathbf{x}_{n+1}\}$, satisfying the above stated conditions, and result is valid for all such sets. \\(3) In the context of the EMD, we come across this situation and find signal specific decomposition, where, we like to obtain signal specific IMFs which are complete, linearly independent (LI) and preserve the energy of signal.

The EPEMD algorithm for the signal $x(t)$, to obtain EPIMFs, is summarized in Algorithm 4:
\\
\begin{tabular}{p{0.99\textwidth}}
\hline
Algorithm 4: Algorithm for EPEMD,  for $i=1,2,\cdots,n$.\\
\hline
${1.}$ Set $x_1(t) = x(t)$.
\newline ${2.}$ Using EMD algorithm, obtain IMF $y_i(t)$ and residue $r_i(t)$ of signal $x_1(t)$.
\newline ${3.}$ Orthogonalize IMF $y_i(t)$ and residue $r_i(t)$ (as explained above to obtain \eqref{st12}) and obtain new orthogonal IMF $c_i(t)$ and residue $c'_{i+1}(t)$.
\newline ${4.}$ Set $x_1(t) = c'_{i+1}(t)$ and repeat steps $2$ to $4$ and end when all the EPIMFs of signal $x(t)$ are obtained.\\
\hline
\end{tabular}
%\section{Gram-Schmidt Orthogonalization Method}
\section{The GSOM, orthogonal EMD and orthogonal IMFs}
In this section, we propose another algorithm, through the GSOM, by exploiting the specific order in which the properties of IMFs and energy of the signal are preserved in the decomposition.
\subsection{The GSOM and Orthogonal EMD}
The GSOM is a process for orthogonalizing a set of signals in an inner product space. Let $Y = \{y_1(t),y_2(t),...,y_n(t)\}$ be a set of $n$ LI signals. A set of orthogonal signals $S = \{s_1(t),s_2(t),...,s_n(t)\}$ is generated from the set $Y$ as follows (for $k=1,2,\dots,n$):
 %$s_k(t)=y_k(t)-\sum_{i=1}^{k-1}c_{ki}s_i(t) \Leftrightarrow$
\begin{equation}
s_k(t)=y_k(t)-\sum_{i=1}^{k-1}c_{ki}s_i(t) \Leftrightarrow
 \left[ \begin{array}{c} y_1(t) \\ y_2(t) \\ \vdots \\ y_n(t) \end{array} \right] = \begin{bmatrix} 1 & 0 & \dots & 0 \\ c_{21} & 1 & \dots & 0 \\ \vdots & \vdots & \ddots & \vdots \\ c_{n1} & c_{n2} & \dots & 1 \end{bmatrix} \left[ \begin{array}{c} s_1(t) \\ s_2(t) \\ \vdots \\ s_n(t) \end{array} \right]\label{gso1}\end{equation}
The $c_{ki}$ is obtained by using inner product $ \langle s_k(t),s_i(t) \rangle =0 \text{, for } k \neq i$, i.e.
%\begin{equation}
$c_{ki}=\int_{0}^Ty_k(t)s_i(t)dt/\int_{0}^Ts_i^2(t) dt$
% \end{equation}
for $i=1,2,\dots,n \text{, and } k\ge i$.
where $T$ is the total observation period of the signals. By taking sum of all the $n$ equations of \eqref{gso1} along with some simple algebraic manipulations, it is shown in \cite{rs3} that
\begin{equation}
\sum_{i=1}^{n}y_i(t)=\sum_{i=1}^{n}c_i{s}_{i}(t) \label{vf1}
\end{equation}
where $c_i=\sum_{k=i}^{n}c_{ki}$ is sum of $i^{th}$ column of the coefficient matrix of \eqref{gso1}.
 It can be easily shown that $c_{ki}=1$, if $k=i$. From set $Y$, there are $n$ choices for selecting first signal, $n-1$ choices for second signal, $n-2$ choices for third signal and 1 choice for last signal, that means there are $n!$ permutations of the set $Y$, and the GSOM would produce $n!$ orthogonal sets of signals from a set of $n$ LI signals. So it looks interesting to explore whether in a particular application, one choice is better than other and also why. We explore this issue in the context of EMD.

In order to ensure the exact orthogonality and to eliminate energy leakage among IMFs, orthogonal EMD (OEMD) based on the GSOM is developed in \cite{rs3}, which generates complete orthogonal IMFs (OIMFs). Let, $Y = \{y_1(t),y_2(t),...,y_n(t)\}$ be a set of $n$ IMFs of the signal $x(t)$ generated from EMD algorithms.
 Through GSOM, OIMFs are obtained from a set of IMFs arranged in order of highest frequency to lowest frequency IMF.
A signal $x(t)$ can be expressed, in terms of IMFs, as \cite{rs3}
\begin{equation}
x(t)= \sum_{i=1}^{n}y_i(t) + r_n(t)=\sum_{i=1}^{n}c_i{s}_{i}(t) + r_n(t)
\end{equation}
An OEMD algorithm produces the residue signal $r_n(t)$ and $n$ orthogonal signals $s_i(t)$, as multiplication of constant $c_i$ on the signal $s_i(t)$ does not affect orthogonality.
\subsection{The OEMD and Orthogonal IMFs}
The OIMFs obtained from OEMD have the following limitations:
(1) Higher mode of oscillation is multiplied by some factor and is being subtracted from lower mode of oscillation in OEMD, which would result in mixing of high frequency components to low frequency one, as shown in simulation results.
% Figure~\ref{fig:EMDFig2}.
(2) Because of the mixing of high frequency components to low frequency one, properties of IMFs are not preserved by signal $s_i(t)$ and hence some of the instantaneous frequencies, derived from the Hilbert transform, becomes negative which has no physical meaning.
(3) Residue signal is not orthogonal to any signal component and hence there is always some energy leakage. As we have shown that the GSOM can generate $n!$ orthogonal sets from a set of $n$ LI signals and OIMFs, generated above, are one such set. We explore other orthogonal set in this section.

To overcome the limitations of OIMFs and to completely stop energy leakage, we propose second EPEMD algorithm which generates two sets of orthogonal IMFs through the GSOM which proceeds in the reverse order (i.e. from residue to first IMF). To obtain the first set of IMFs, we propose to apply the GSOM starting from the residue and finally reaching to the first IMF and we refer it as reverse orthogonal IMFs (ROIMFs). The second set of IMFs are obtained by following the same order (i.e. the reverse order), with only difference being that all IMFs and residue are made zero mean before applying the GSOM, and we refer it as reverse orthogonal and uncorrelated IMFs (ROUIMFs).

If vectors of zero mean are orthogonal or uncorrelated, then they are `orthogonal and uncorrelated' (i.e. orthogonality and uncorrelatedness is the same) \cite{rs11}, and the LI, orthogonal and uncorrelated subspaces, of inner product space, are shown in Figure~\ref{fig:Vspace}.
%Let $\mathbf{X}$ and $\mathbf{Y}$ be vectors in inner product space. Then
%(1) $\mathbf{X}$ and $\mathbf{Y}$ are LI if and only if $a\mathbf{X}+b\mathbf{X}=0$; when $a=0, b=0$.
%(2) $\mathbf{X}$ and $\mathbf{Y}$ are orthogonal if and only if $\mathbf{X^TY}=0$.
%(3) $\mathbf{X}$ and $\mathbf{Y}$ are uncorrelated if and only if $\mathbf{(X-I_1}X_{mean})^T\mathbf{(Y-I_1}Y_{mean})=0$, where $\mathbf{I_1}$ is a vector of %ones \cite{rs11}. From (2) and (3) it is clear that if vectors of zero mean are orthogonal or uncorrelated, then they are `orthogonal and uncorrelated'.

The ROUIMFs are obtained through the GSOM from a set of mean removed residue and IMFs which are arranged in order of lowest frequency to highest frequency components (i.e. $\{r'_n(t), y'_n(t),y'_{n-1}(t),...,y'_1(t)\}$, with $r'_n(t)=[r_n(t)-r_{n,mean}]$ and $y'_i(t)= [y_i(t)-y_{i,mean}]$, where mean of the a signal $g(t)$ is defined as $\frac{1}{T}\int_{0}^{T}g(t)\,dt$). Thus, lower mode of oscillations is multiplied by some coefficient and subtracted from higher mode of oscillations in the GSOM to produces IMFs that preserve the properties of IMFs. It is also verified through large number of simulations that, with this order of orthogonalization, the properties of IMFs are preserved.
Let $z_1(t)=r'_n(t), z_2(t)=y'_{n}(t), \cdots$, and $z_{n+1}(t)=y'_1(t)$.
A signal $x(t)$ is decomposed in ${n+2}$ orthogonal components without any energy leakage as follows:
\begin{equation}
x(t)=\sum_{i=1}^{n+1}z_i(t)+C \label{OUIMFsEq1}
\end{equation}
where constant $C=\sum_{i=1}^{n}y_{i,mean}+r_{n,mean}$ is mean of the signal $x(t)$. By applying the GSOM on $\sum_{i=1}^{n+1}z_i(t)$, as in \eqref{vf1}, we obtain
\begin{equation}
\sum_{i=1}^{n+1}z_i(t)=\sum_{i=1}^{n+1}c_i{s}_{i}(t)=\sum_{i=1}^{n+1}{p}_{i}(t)
\end{equation}
The energy of the signal $x(t)$ can be easily seen to be:
\begin{equation}
\int_{0}^Tx^2(t) dt=\int_{0}^T\left(\sum_{i=1}^{n+1}p^2_i(t)+C^2\right) dt
\end{equation}
This procedure of the GSOM produces a constant signal along with ${n+1}$ orthogonal and uncorrelated components of the signal (which includes residue signal as well). There are following benefits of deriving ROUIMFs as compared to other (OIMFs and FOUIMFs, etc.) set of IMFs:
%\\(1) In the GSOM, lower mode of oscillations are being partially subtracted from higher mode of oscillations,
% smoothness is maintained in the ROUIMF's as present in IMF's.
\\(1) The ROUIMFs obtained by applying GSOM on IMFs preserve the properties of IMFs, whereas most of the FOUIMF's are not able to maintain the properties of IMFs.
%(3) There is redistribution of energy in process of obtaning ROUIMF's and  IMF's of lower mode of oscillations are having more energy which is useful in cases %where low frequency components signals are important to monitor and observe.
(2) The residue signal is also orthogonal and uncorrelated to all other IMFs and therefore, there is no energy leakage.
\\This process can be easily extended to obtain the orthogonal MEMD (OMEMD) from the MEMD, and we can derive orthogonal multivariate IMFs.
\\
\begin{tabular}{@{} l @{}}
\hline
We use following notations:\\
\hline
$\{x(t)\} \to\text{EMD}\mapsto \{y_1(t),...,y_n(t),r_n(t)\} \text{=\{IMFs,residue\}}$. \\
$\{x(t)\} \to\text{EPEMD}\mapsto \{c_1(t),...,c_n(t),c_{n+1}(t)\} \text{=\{EPIMFs\}}$. \\
$\{y_1(t),y_2(t),\dots,y_n(t)\} \to\text{GSOM}\mapsto \text{ \{OIMFs\} }$. \\
$\{y_1(t),\dots,y_n(t),r_n(t)\} \to\text{GSOM}\mapsto  \text{ \{FOIMFs\}}$. \\
$\{r_n(t),y_n(t),\dots,y_1(t)\} \to\text{GSOM}\mapsto \text{ \{ROIMFs\}}$. \\
$\{y'_1(t),\dots,y'_n(t),r'_n(t)\} \to\text{GSOM}\mapsto  \text{ \{FOUIMFs\}}$. \\
$\{r'_n(t),y'_n(t),\dots,y'_1(t)\} \to\text{GSOM}\mapsto \text{ \{ROUIMFs\}}$. \\
$\{x_1(t),\dots,x_m(t)\} \to\text{MEMD}\mapsto\{y_{11}(t),\dots,r_{1n}(t)\},\dots,\{y_{m1}(t),\dots,r_{mn}(t)\}$. \\
$\{r_{1n}(t),...,y_{11}(t)\},...,\{r_{mn}(t),...,y_{m1}(t)\} \to\text{GSOM}\mapsto \text{\{ROIMFs\},...,\{ROIMFs\}}$.\\
\hline
\end{tabular}
\section{Simulation results}
The online available MATLAB software for EMD and EEMD \cite{rs23}, and for MEMD \cite{rs24} have been used in simulation results. The objectives of the simulations are (1) to calculate and compare energy leakage, percentage energy error and index of orthogonality from the proposed algorithms as well as from the EMD, EEMD and MEMD algorithms, by using the simulated and real file signals. (2) to use the proposed algorithm for time-frequency analysis of a chirp signal and compare result with the EMD and EEMD. (3) to test statistical significance of IMFs generated by the proposed algorithms.
\subsection{The comparison of energy leakage between EMD and EPEMD}
\begin{figure}[!t]
\centering
\includegraphics[angle=0,width=0.99\textwidth]{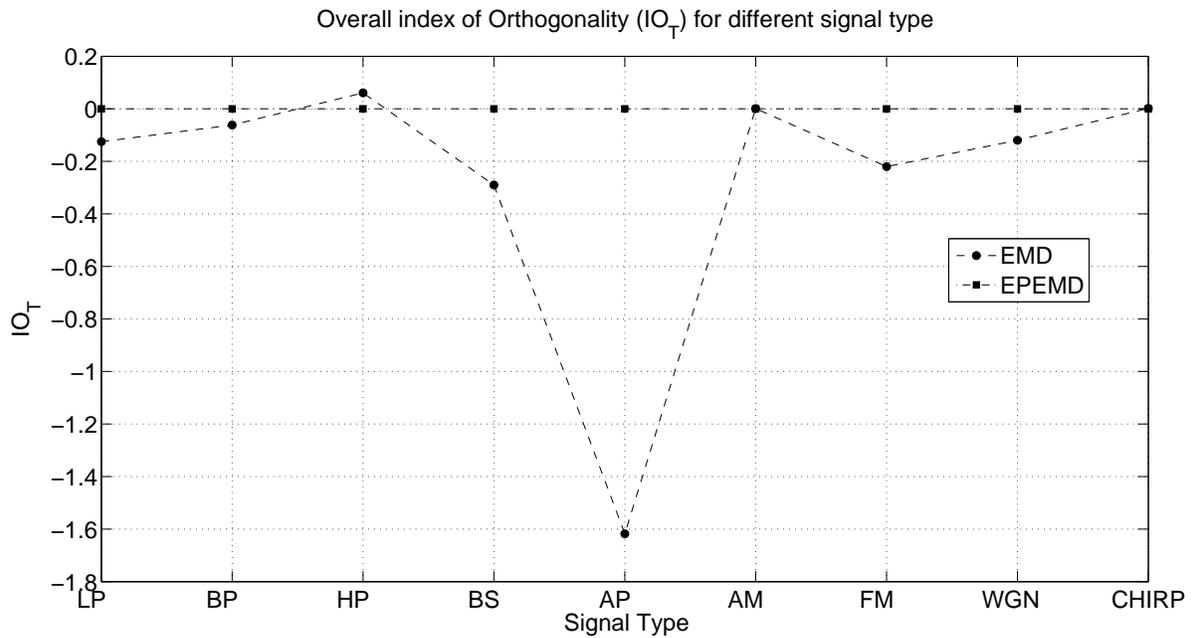}
\caption{Overall Index of Orthogonality for different type of signal}
\label{fig:OIt1}
\end{figure}

\begin{figure}[!t]
\centering
\includegraphics[angle=0,width=0.99\textwidth]{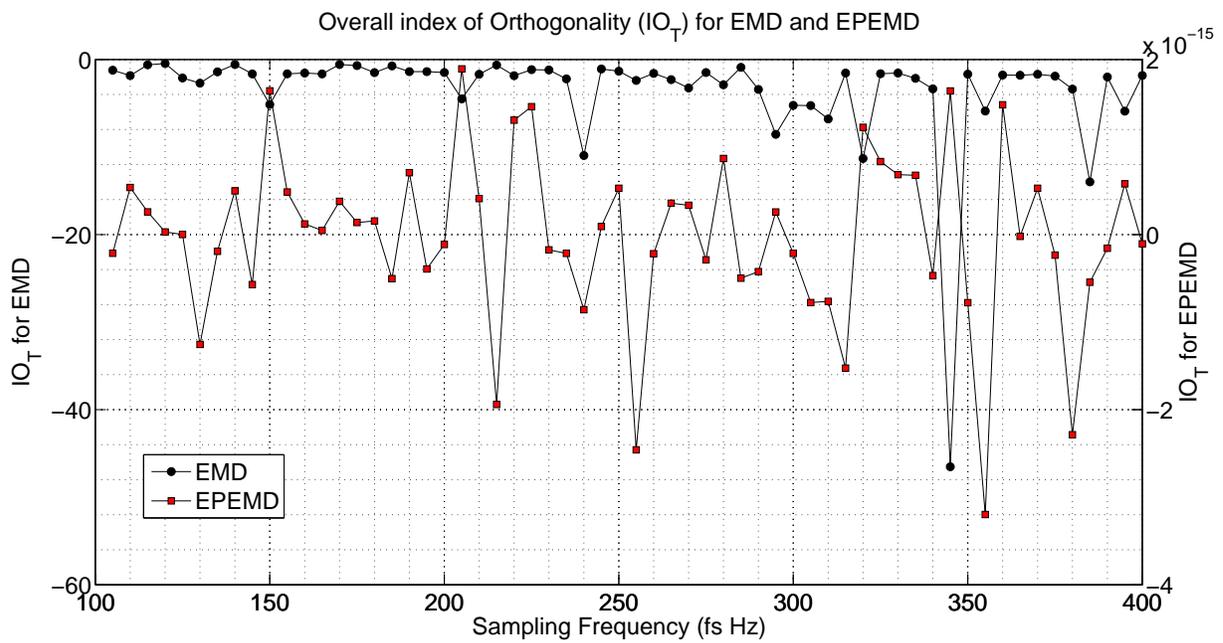}
\caption{$IO_T$ for different sampling frequency}
\label{fig:OIt2}
\end{figure}

\begin{figure}[!t]
\centering
\includegraphics[angle=0,width=0.99\textwidth]{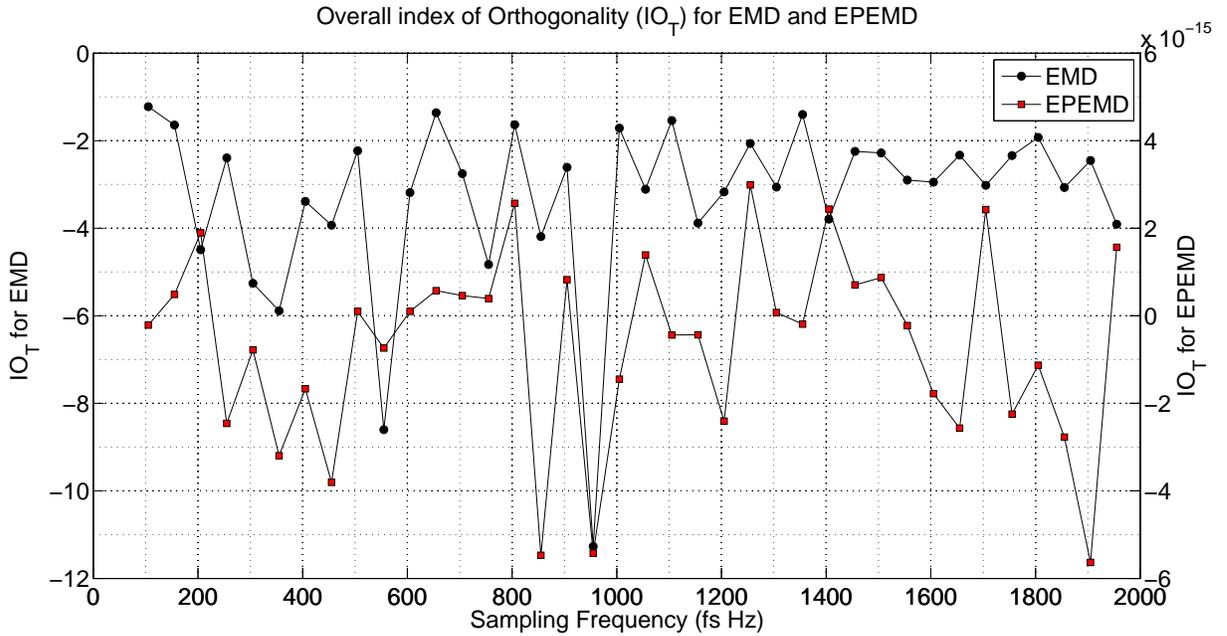}
\caption{$IO_T$ for different sampling frequency}
\label{fig:OIt3}
\end{figure}

The overall index of orthogonality ($IO_T$) is shown in Figure~\ref{fig:OIt1} for the following different type of signals, with $A_1$=100, $A_2$=1, sampling frequency $F_s$=150 and time duration of 0 to 10 sec.
(1) Low pass signal, $LP=\sum_{i=1}^{20}[A_2 sin(2\pi(50-i)t)+A_1 sin(2\pi(1+i)t)]$;
(2) Band pass signal, $BP=\sum_{i=1}^{20}[A_2 sin(2\pi (50-i)t)+A_1 sin(2\pi(15+i)t)+A_2 sin(2\pi(1+i)t)]$;
(3) High pass signal, $HP=\sum_{i=1}^{20}[A_1 sin(2\pi(50-i)t)+A_2 sin(2\pi(1+i)t)]$;
(4) Band stop signal, $BS=\sum_{i=1}^{20}[A_1 sin(2\pi(50-i)t)+A_2 sin(2\pi(15+i)t)+A_1 sin(2\pi(0+i)t)]$;
(5) All pass signal, $AP=\sum_{i=1}^{50}[A_1 sin(2\pi it)]$;
(6) AM signal, $AM=(1+A_2 sin(2\pi3t))\cdot(A_1 sin(2\pi20t))$;
(7) FM signal, $FM=A_1 sin((2\pi10+5 sin(2\pi3t))t)$;
(8) White Gaussian noise, WGN (normal distribution with mean 0 and standard deviation 1); and
(9) CHIRP (linear chirp of amplitude 100 and frequency 0.1 Hz to 50 Hz).
Figure~\ref{fig:OIt2} and Figure~\ref{fig:OIt3} show $IO_T$ for different sampling rates, generated from signal $s(t)=\sum_{f_i=1}^{50}[100\cdot sin(2\pi f_it)]$ with time duration of 0 to 10 sec, sampling rate $F_s$=105 to 400 Hz with increment of 5 Hz and $F_s$=105 to 2000 Hz with increment of 50 Hz, respectively.
As shown in these simulation, $IO_T$ is varying with type of signal and sampling frequency and sometimes it is very high (e.g. in Figure~\ref{fig:OIt2} peak value of $IO_T\approx-58$) in EMD whereas in case of EPEMD $IO_T$ is always in the range of $10^{-15}$ which is almost zero.
\subsection{Real life time series decomposition}
In this simulation we take real life time series to compare the energy leakage and percentage energy error among the IMFs generated by the EMD, EEMD and OEMD algorithms.
\subsubsection{The annual mean global surface temperature anomaly time series analysis}
\begin{figure}[!t]
\centering
\includegraphics[angle=0,width=0.99\textwidth]{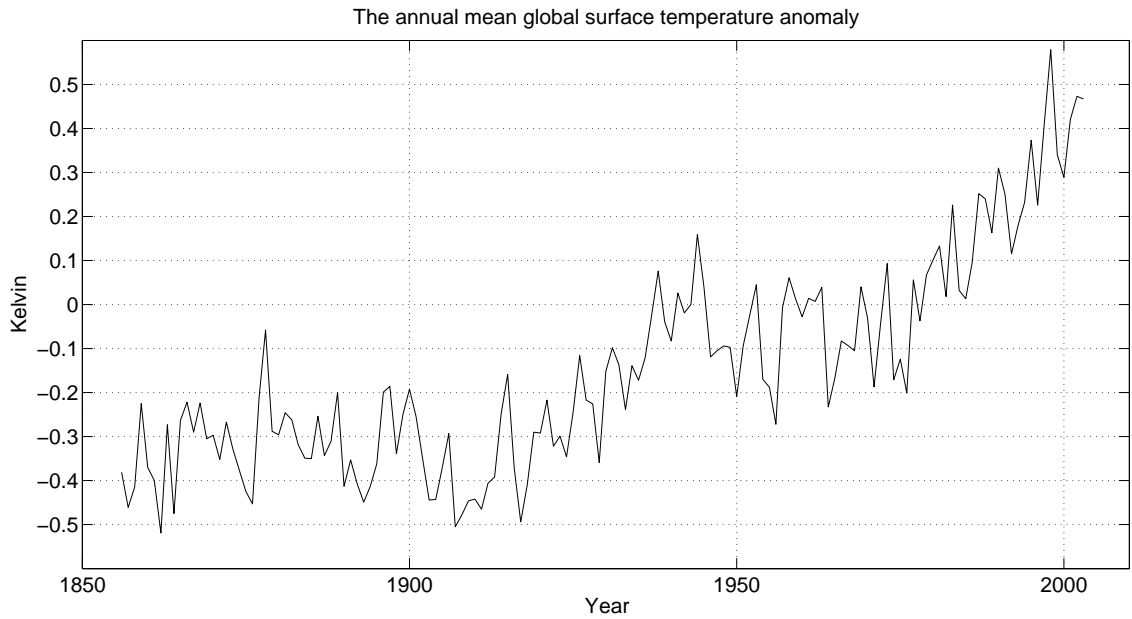}
\caption{Annual mean global surface temperature anomaly}
\label{fig:Annualmean}
\end{figure}

\begin{figure}[!t]
\centering
\includegraphics[angle=0,width=0.99\textwidth]{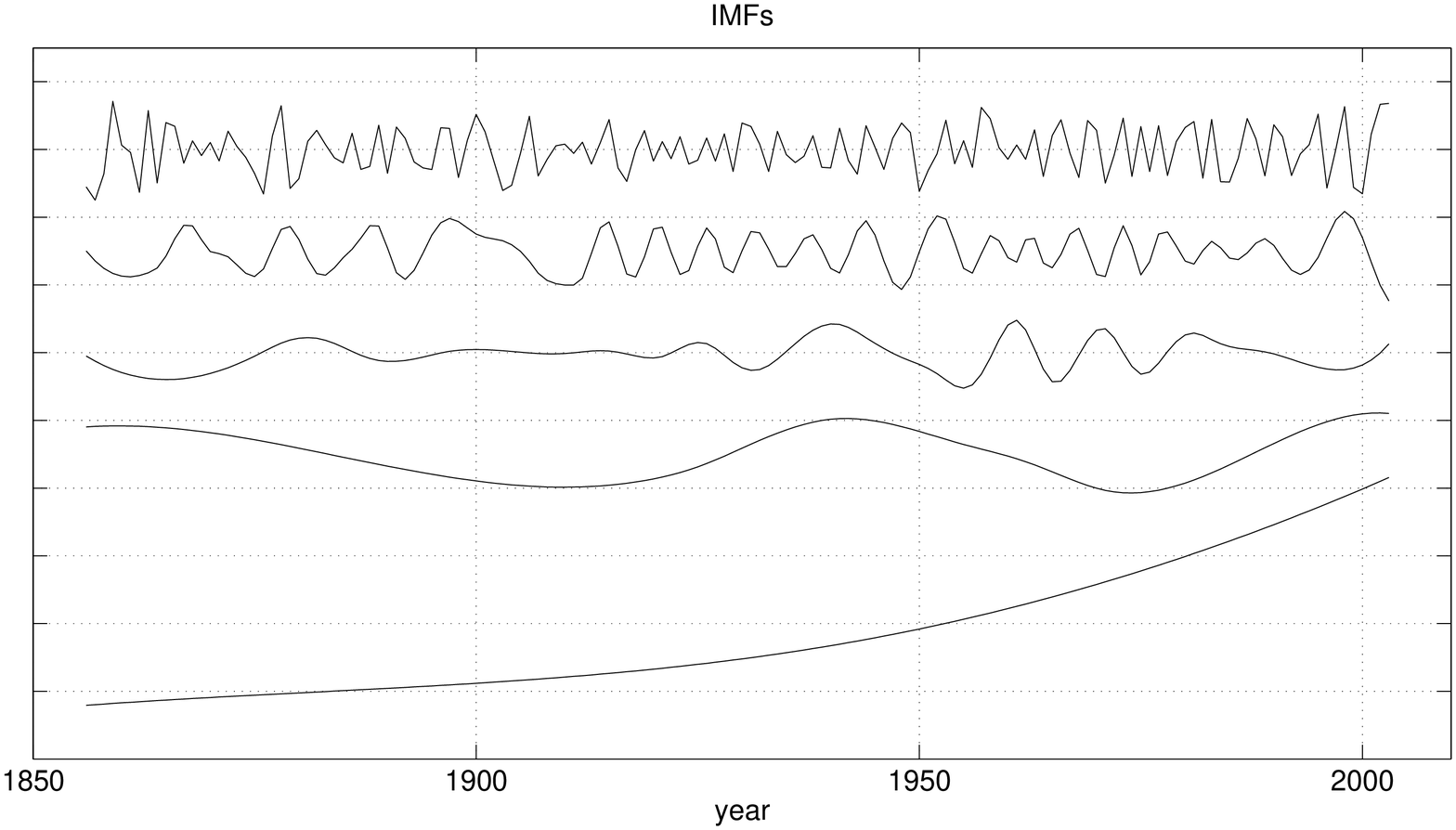}
\captionof{figure}{IMFs $y_1$ to $y_4$ and $(y_5+y_6+r_6)$ obtained from EMD.}
\label{fig:EmdFig1}
\end{figure}

\begin{figure}[!t]
\centering
\includegraphics[angle=0,width=0.99\textwidth]{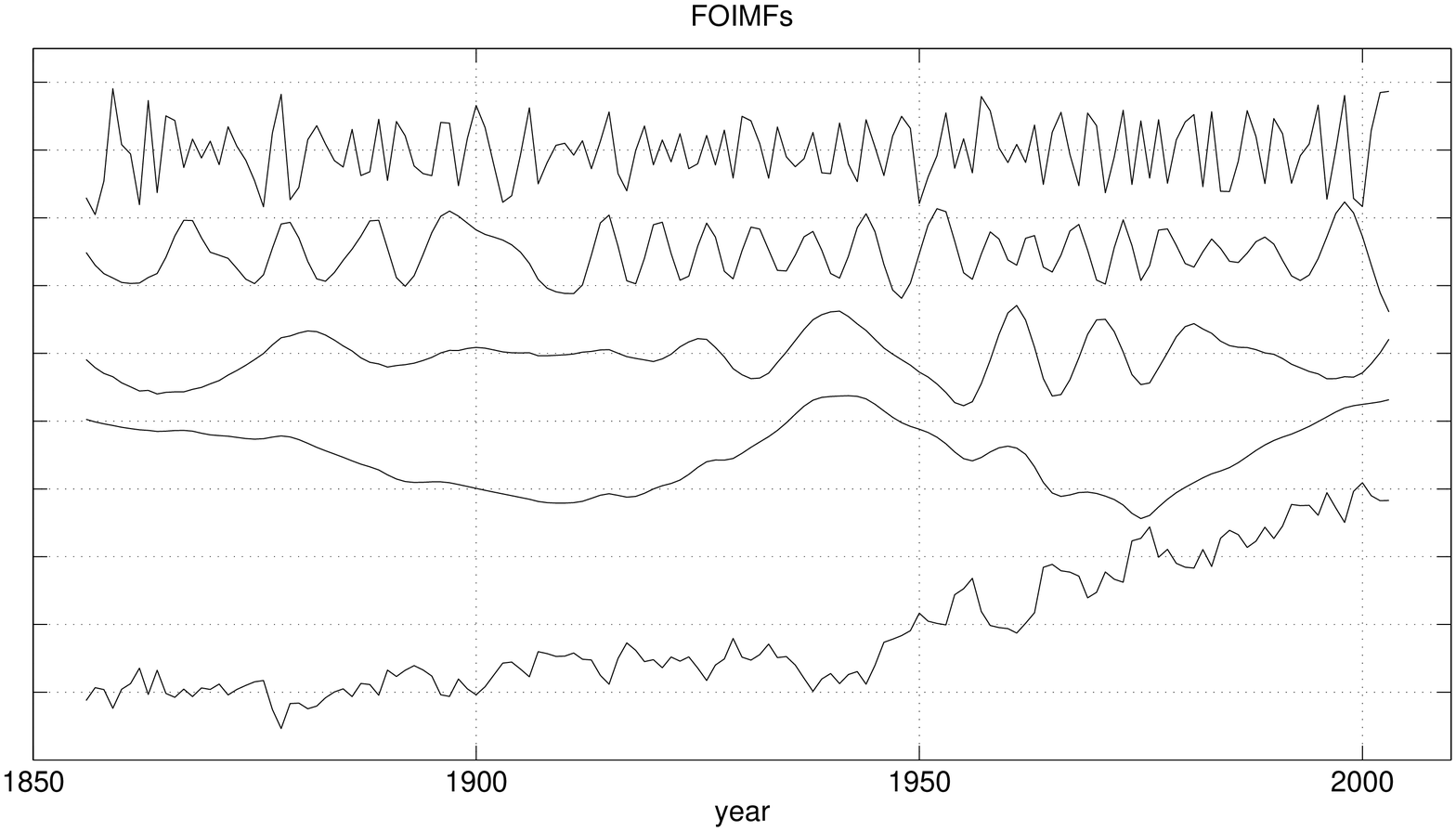}
\captionof{figure}{FOIMFs $y_1$ to $y_4$ and $(y_5+y_6+r_6)$ obtained from EMD. High frequency components are mixed in low frequency ones.}
\label{fig:EmdFig2}
\end{figure}

\begin{figure}[!t]
\centering
\includegraphics[angle=0,width=0.99\textwidth]{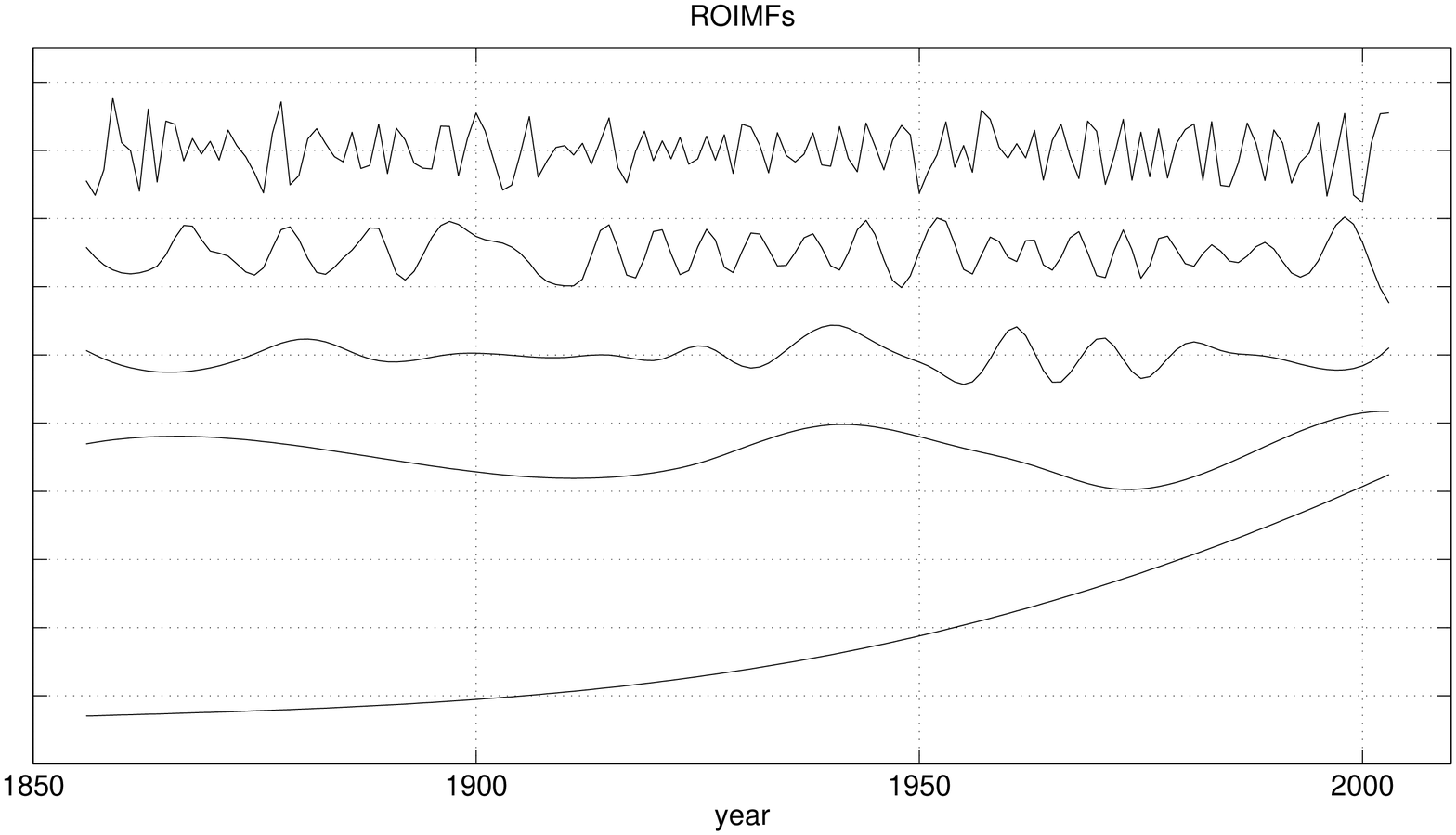}
\captionof{figure}{ROIMFs $y_1$ to $y_4$ and $(y_5+y_6+r_6)$ obtained from EMD}
\label{fig:EmdFig3}
\end{figure}

\begin{figure}[!t]
\centering
\includegraphics[angle=0,width=0.99\textwidth]{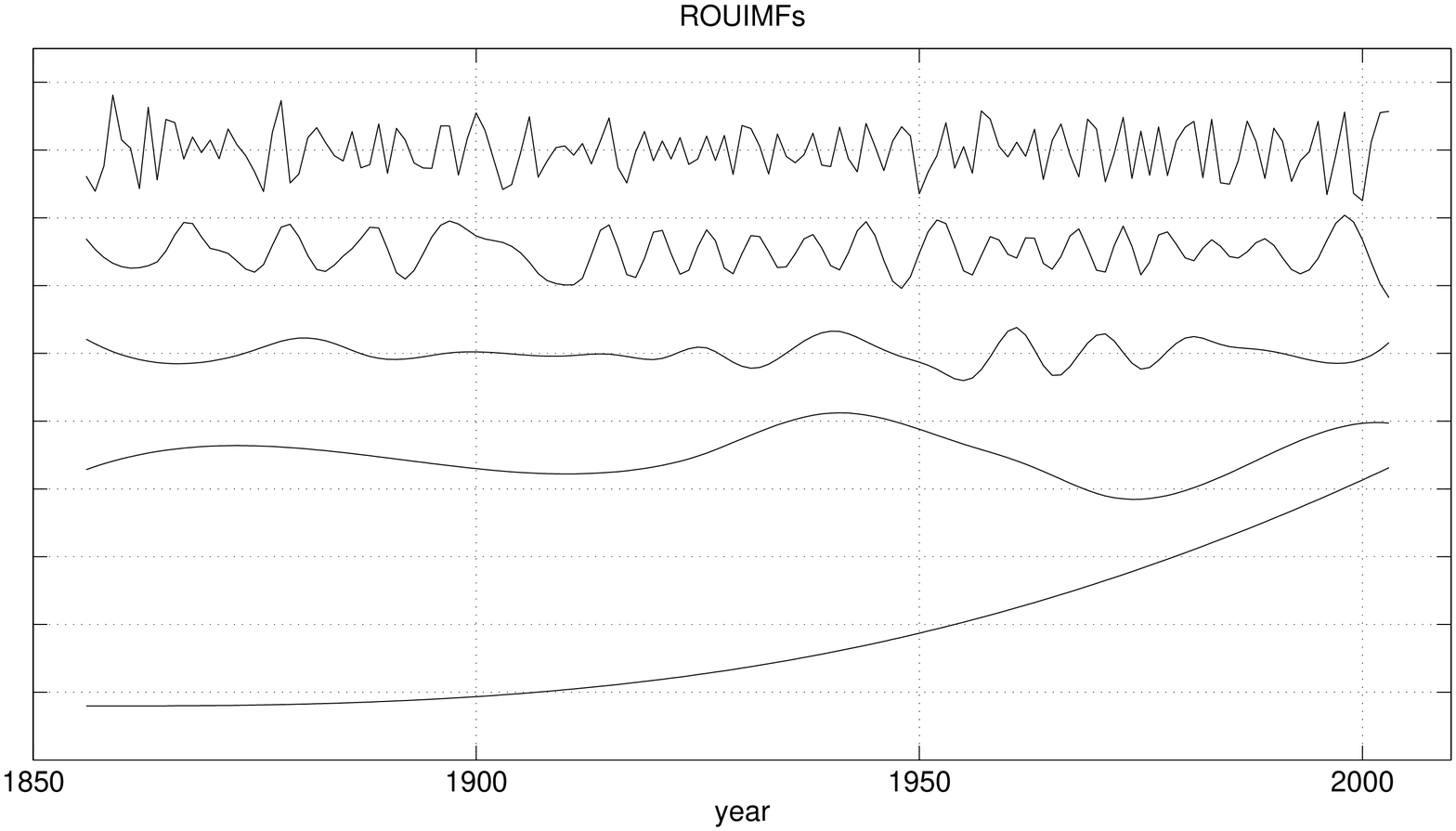}
\captionof{figure}{ROUIMFs $y_1$ to $y_4$ and $(y_5+y_6+r_6)$ plus DC component obtained from EMD}
\label{fig:EmdFig31}
\end{figure}

\begin{figure}[!t]
\centering
\includegraphics[angle=0,width=0.99\textwidth]{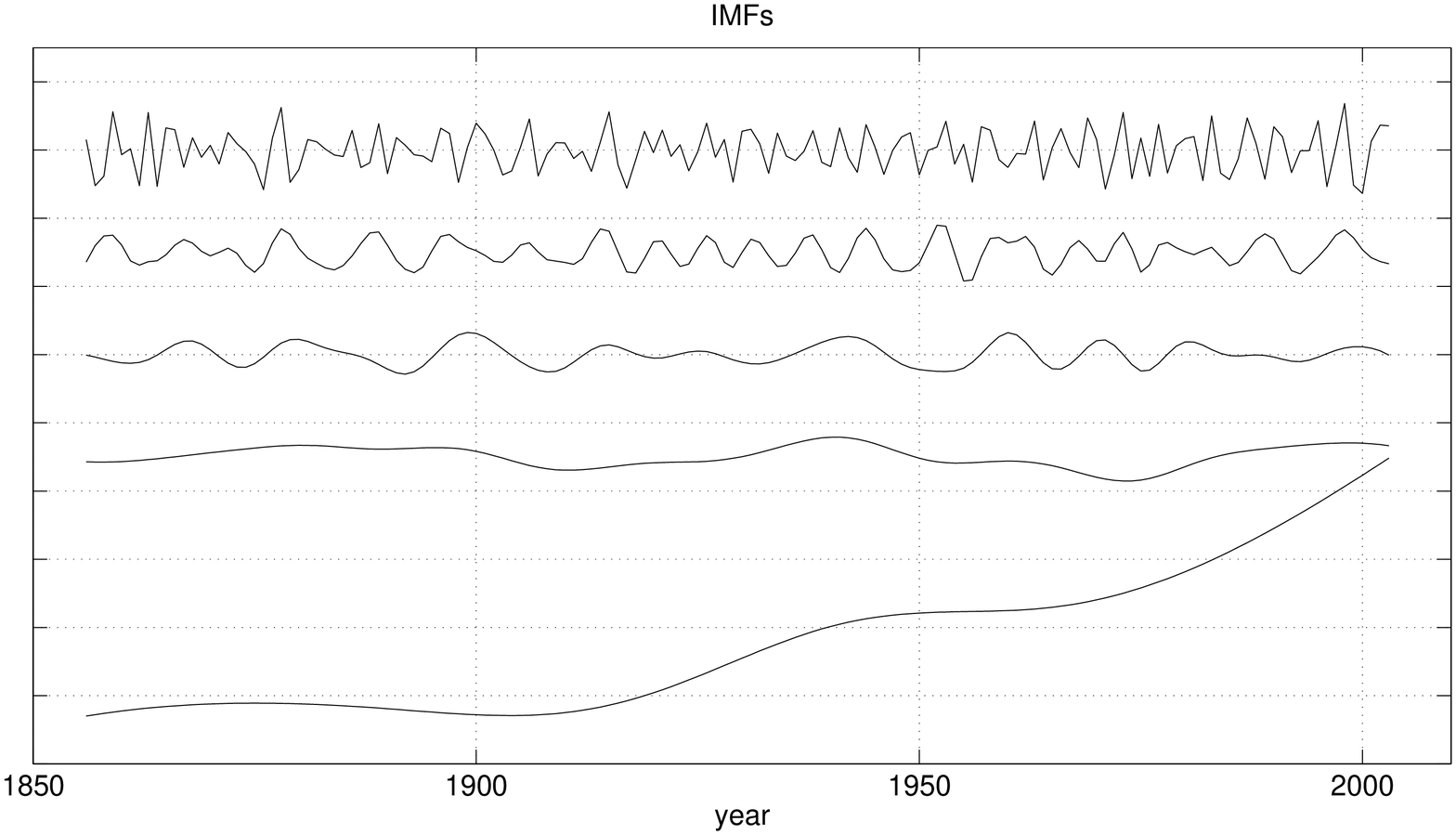}
\captionof{figure}{IMFs $y_1$ to $y_4$ and $(y_5+y_6+r_6)$ obtained from EEMD}
\label{fig:EmdFig4}
\end{figure}

\begin{figure}[!t]
\centering
\includegraphics[angle=0,width=0.99\textwidth]{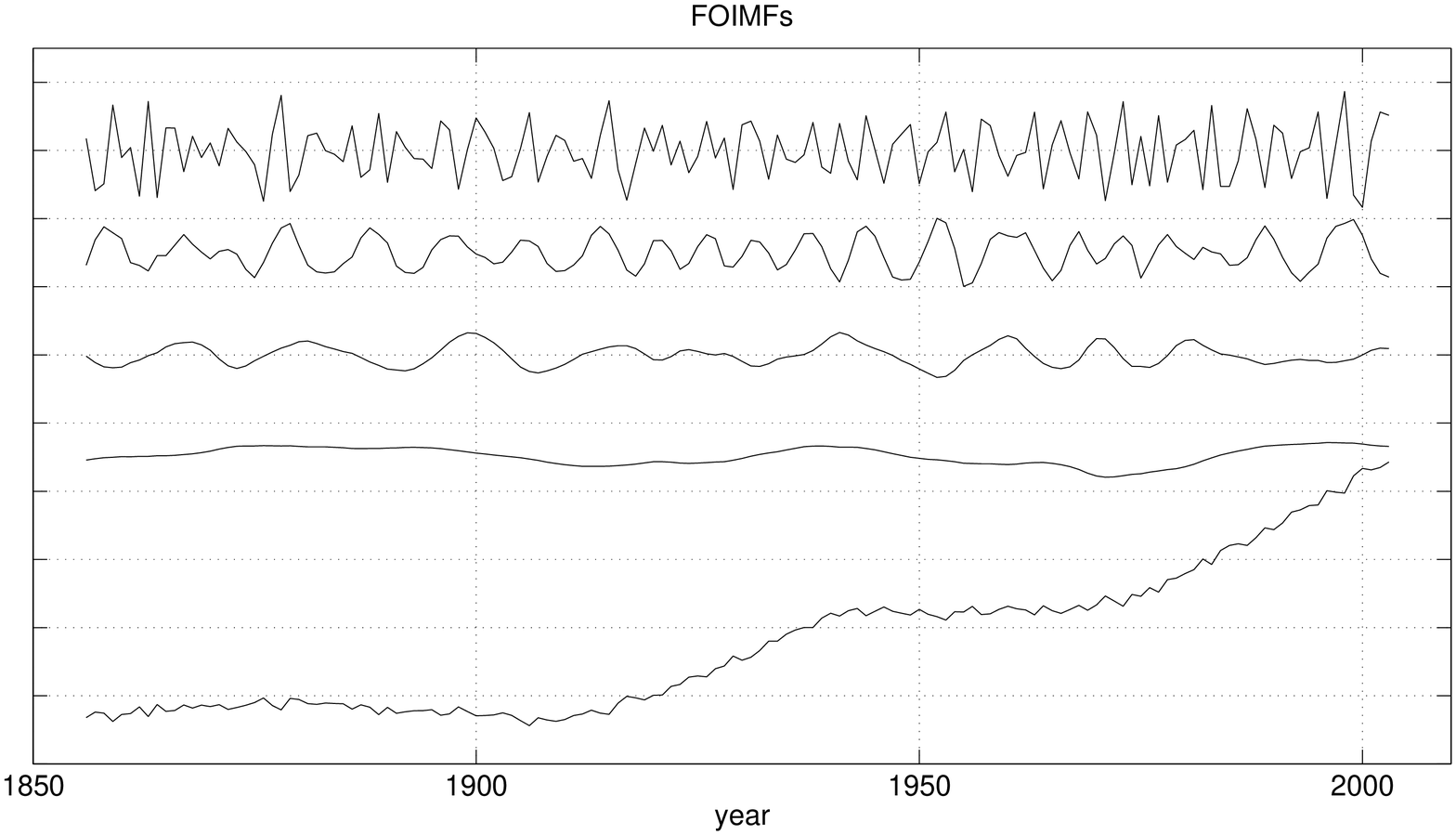}
\captionof{figure}{FOIMFs $y_1$ to $y_4$ and $(y_5+y_6+r_6)$ obtained from EEMD. High frequency components are mixed in low frequency ones.}
\label{fig:EmdFig5}
\end{figure}

\begin{figure}[!t]
\centering
\includegraphics[angle=0,width=0.99\textwidth]{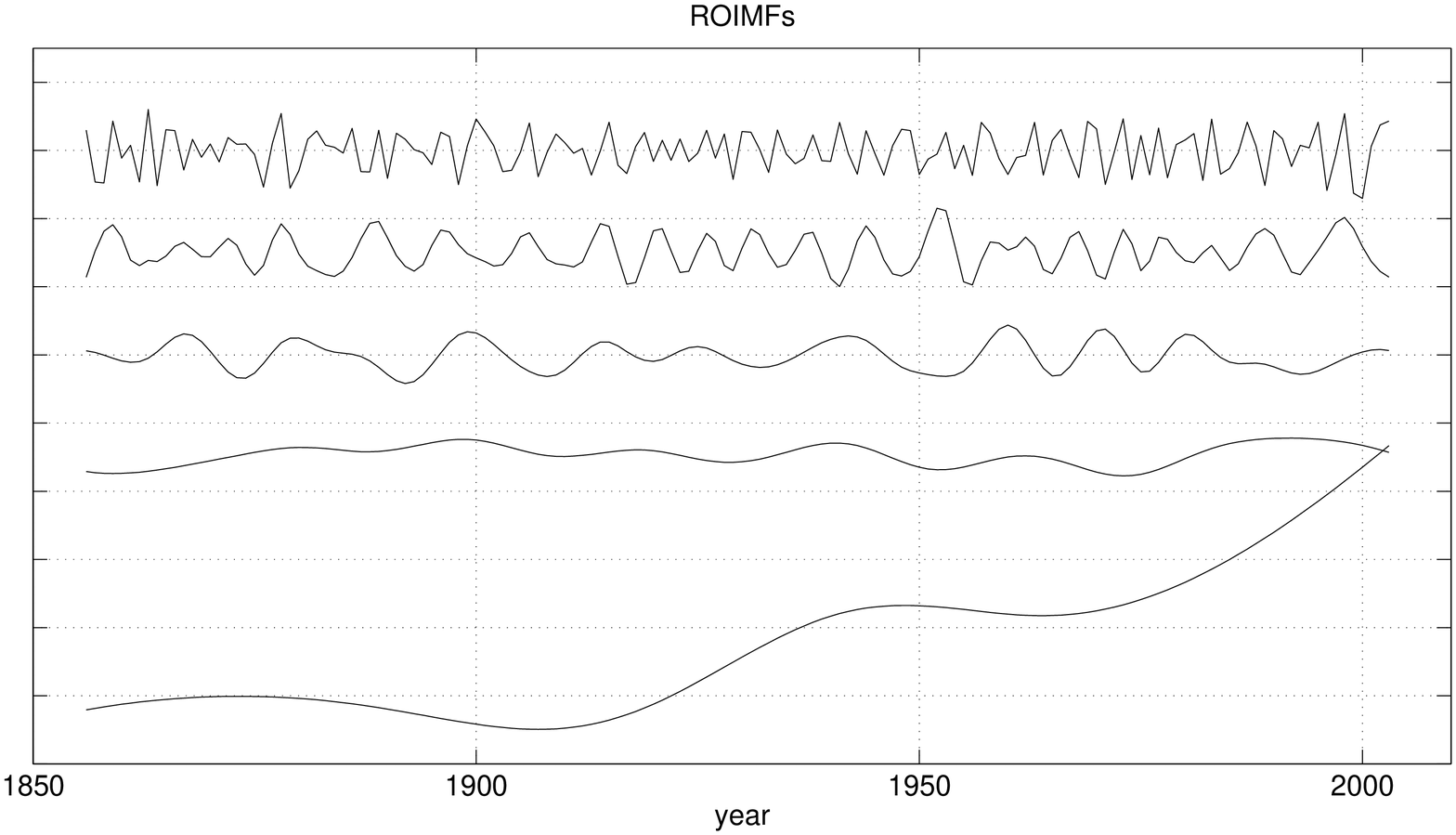}
\captionof{figure}{ROIMFs $y_1$ to $y_4$ and $(y_5+y_6+r_6)$ obtained from EEMD}
\label{fig:EmdFig6}
\end{figure}

\begin{figure}[!t]
\centering
\includegraphics[angle=0,width=0.99\textwidth]{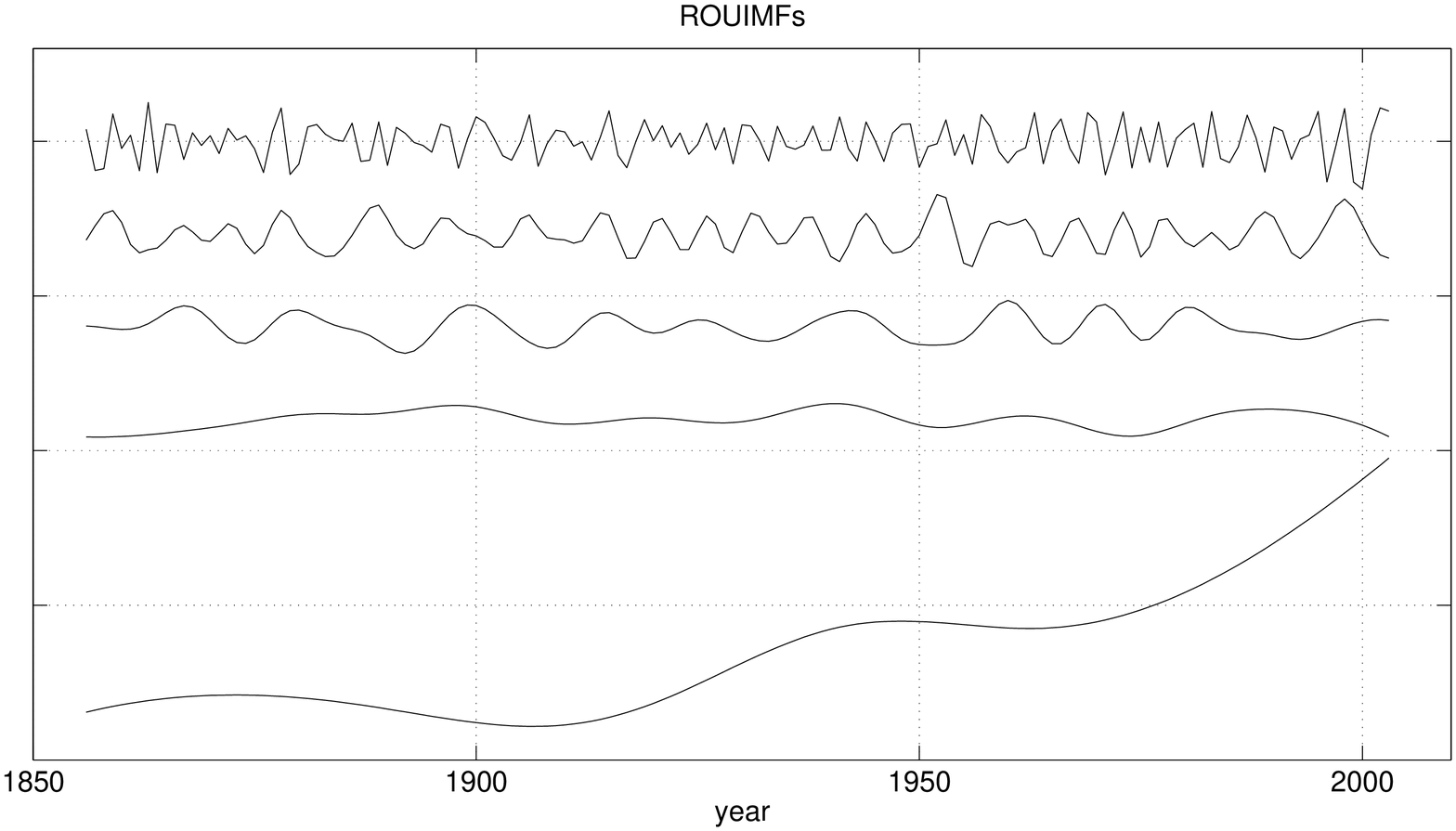}
\captionof{figure}{ROUIMFs $y_1$ to $y_4$ and $(y_5+y_6+r_6)$ plus DC component obtained from EEMD}
\label{fig:EmdFig61}
\end{figure}

In order to compare and demonstrate the advantage of the OEMD with ROIMFs, we applied the proposed method to data ``the annual mean global surface temperature anomaly", online available~\cite{rs23}, as shown in Figure~\ref{fig:Annualmean}. %Data and matlab code of EEMD algorithm used in these simulations has been taken from %\url{http://rcada.ncu.edu.tw/research1_clip_ex.htm}.
The sets of IMFs, FOIMFs, ROIMFs and ROUIMFs obtained by the decomposition of the data through the EMD and EEMD methods are shown in Figures~\ref{fig:EmdFig1} to~\ref{fig:EmdFig61}, which shows significant improvements in ROIMFs and ROUIMFs as compare to FOIMFs. The overall Index of Orthogonality $IO_T$ for EMD and EEMD are given in Table ~\ref{table:IOO}, which clearly indicates better performance of ROIMFs over others.
\begin{table}[!t]
\caption{The overall Index of Orthogonality $IO_T$} % title of Table
\centering % used for centering table
\begin{tabular}{l l l l l} % centered columns (4 columns)
\hline %inserts double horizontal lines
&IMFs &OIMFs &FOIMFs &ROIMFs \\  % inserts table
\hline % inserts single horizontal line
EMD & 48.5e-3 & 79.2e-3 & 92.3e-17 & -41.9e-18\\
EEMD & -15.1e-3 & -69.7e-3 & 11.5e-17 & -56.0e-18\\ % [1ex] adds vertical space
\hline %inserts single line
\end{tabular}
\label{table:IOO} % is used to refer this table in the text
\end{table}
\begin{table}[!t]
\caption{The signal energy $E_x=11.3752$, components and sum of components energies and $Pee$ for IMFs, OIMFs, FOIMFs and ROIMFs obtained from EMD. $E^*_7$ is energy of residue component.}
\centering % used for centering table
\begin{tabular}{l l l l l} % centered columns (4 columns)
\hline %inserts double horizontal lines
&IMFs &OIMFs &FOIMFs &ROIMFs \\  % inserts table
\hline % inserts single horizontal line
$E_1$ & 707.3e-3 & 632.1e-3 & 1.15 & 694.7e-3\\
$E_2$ & 475.2e-3 & 407.2e-3 & 718.53e-3 & 429.6e-3\\ % [1ex] adds vertical space
$E_3$ & 277.05e-3 & 115.3e-3 & 601.6e-3 & 195.1e-3\\
$E_4$ & 783.5e-3 & 739.2e-3 & 1.4 & 552.2e-3\\
$E_5$ & 88.5e-3 & 88.3e-3 & 660.52e-3 & 12.7e-3\\
$E_6$ & 2.5e-6 & 2.1e-6 & 5.0 & 73.8e-3\\
$E^*_7$ & 8.5 & 8.5 & 1.8 & 9.4\\
$E_T$ & 10.8 & 10.5 & 11.4 & 11.4\\
$Pee$ & 4.85 & 7.9 & 124.9e-15 & 46.8e-15\\
\hline %inserts single line
\end{tabular}
\label{table:emdEr} % is used to refer this table in the text
\end{table}
The value of partial orthogonality index of IMFs are in the range of $10^{-3}$ and for FOIMFs and ROIMFs are in the range of $10^{-18}$. The index of total and partial orthogonality, and $Pee$ are least in ROIMFs. The value of partial orthogonality index and $Pee$ for ROUIMFs is almost the same as the value of ROIMFs.
\subsubsection{The Elcentro Earthquake May 18, 1940 North-South Component time series analysis}
\begin{figure}[!t]
\centering
\includegraphics[angle=0,width=0.99\textwidth]{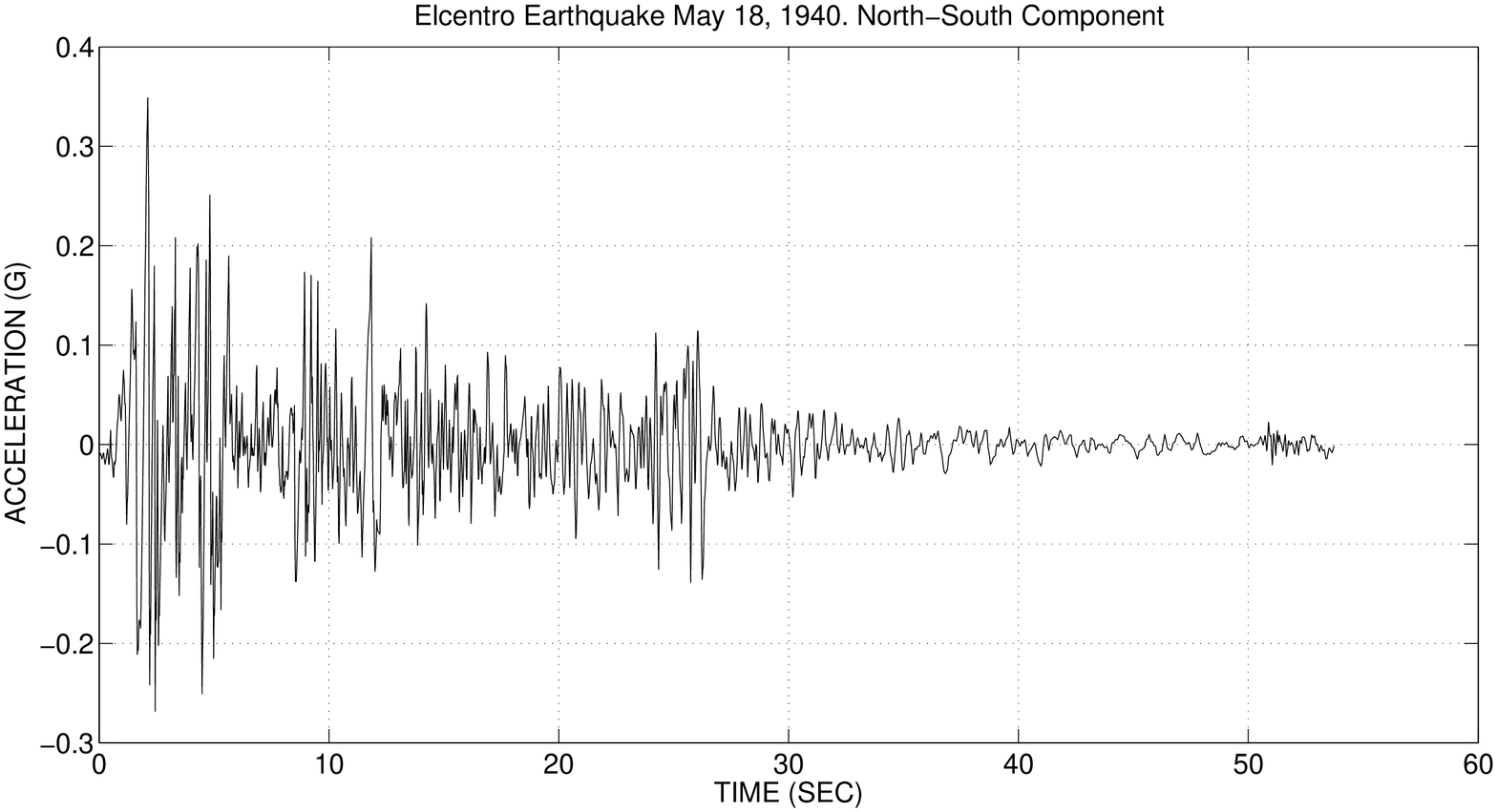}
\caption{Elcentro Earthquake May 18, 1940 North-South Component}
\label{fig:ElcentroEarthquake}
\end{figure}

\begin{figure}[!t]
\centering
\includegraphics[angle=0,width=0.99\textwidth]{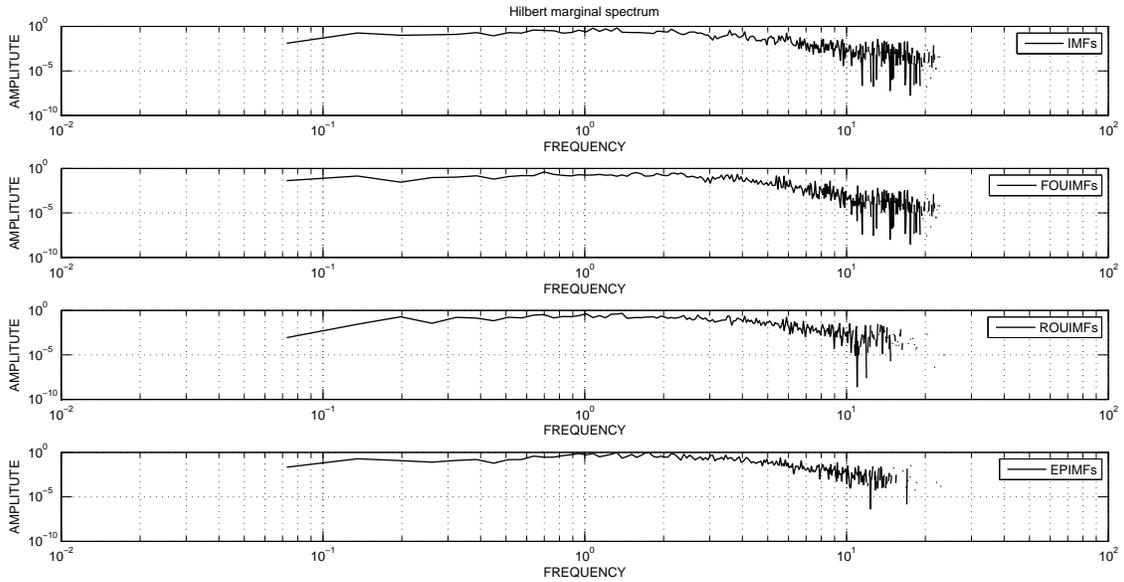}
\caption{Hilbert marginal spectrum for Elcentro Earthquake May 18, 1940 North-South Component time series }
\label{fig:HlbrtMS}
\end{figure}

The Elcentro Earthquake data has been taken from~\cite{rs27} and is shown in Figure~\ref{fig:ElcentroEarthquake}. The percentage errors in total signal energies ($Pee$) for IMFs, OIMFs, and ROIMFs obtained from EMD, for the Elcentro Earthquake time series data, are given in Table \ref{table:EeqIOO}, which indicates better performance (i.e. least value of $Pee$) of ROIMFs over others.
The Hilbert marginal spectrum for IMFs, FOUIMFs, ROUIMFs and EPIMFs derived through EMD of the ElCentro earthquake time series data is shown in Figure \ref{fig:HlbrtMS}. As there are no $Pee$ and no energy leakage among ROUIMFs, and no $Pee$ in EPIMFs, therefore, the Hilbert marginal spectrum of ROUIMFs and EPIMFs can more accurately and faithfully characterize the signal energy distribution at each frequency components.
\begin{table}[!t]
\caption{$Pee$ for IMFs, OIMFs, and ROIMFs obtained from EMD for Elcentro Earthquake data.}
\centering % used for centering table
\begin{tabular}{l l l l} % centered columns (4 columns)
\hline %inserts double horizontal lines
&IMFs &OIMFs &ROIMFs \\  % inserts table
\hline % inserts single horizontal line
$Pee$ & -26 &-2.3e-2 & 5.55e-13 \\
\hline %inserts single line
\end{tabular}
\label{table:EeqIOO} % is used to refer this table in the text
\end{table}
\subsection{The comparison of percentage energy error (Pee) between the MEMD and OMEMD}
\begin{figure}[!t]
\centering
\includegraphics[angle=0,width=0.99\textwidth]{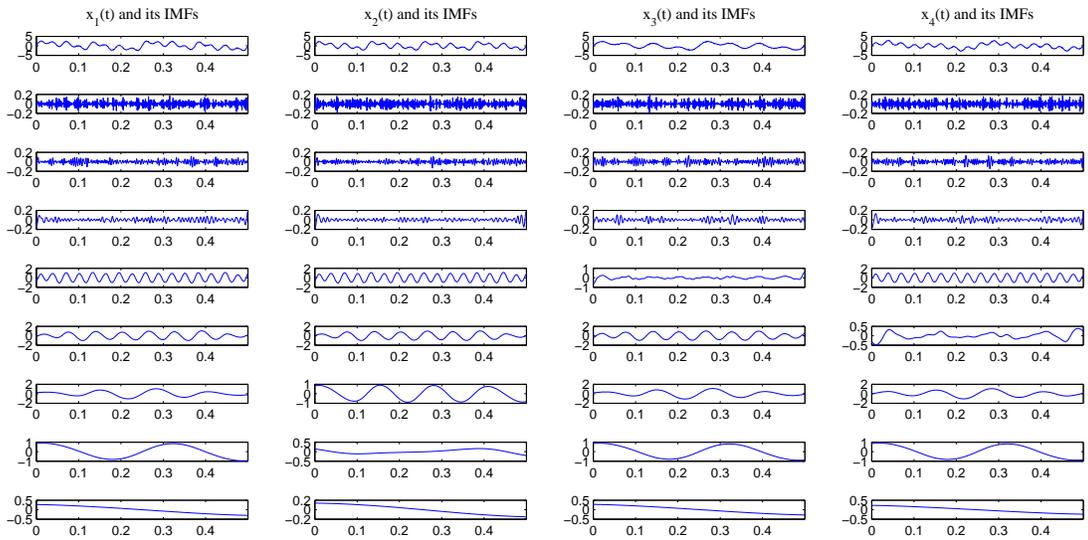}
\caption{4-variate signal $x_j(t)$ (first row) and its IMFs (second row onwards) obtained from MEMD of~\eqref{exam10} in each column.}
\label{fig:memdImfs}
\end{figure}

\begin{figure}[!t]
\centering
\includegraphics[angle=0,width=0.99\textwidth]{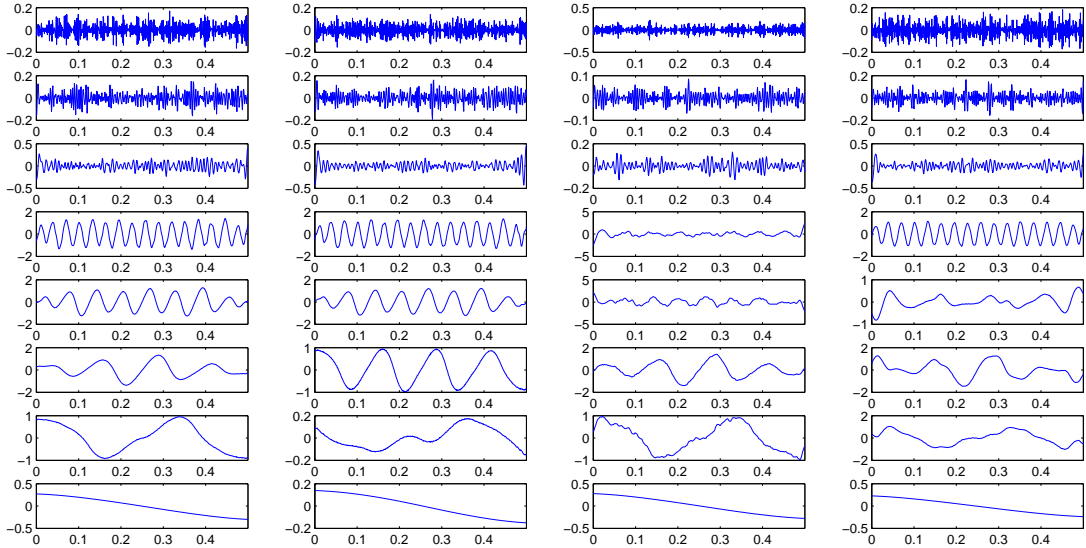}
\caption{FOIMFs obtained from MEMD. High frequency components are mixed in low frequency ones.}
\label{fig:memdFoImfs}
\end{figure}

\begin{figure}[!t]
\centering
\includegraphics[angle=0,width=0.99\textwidth]{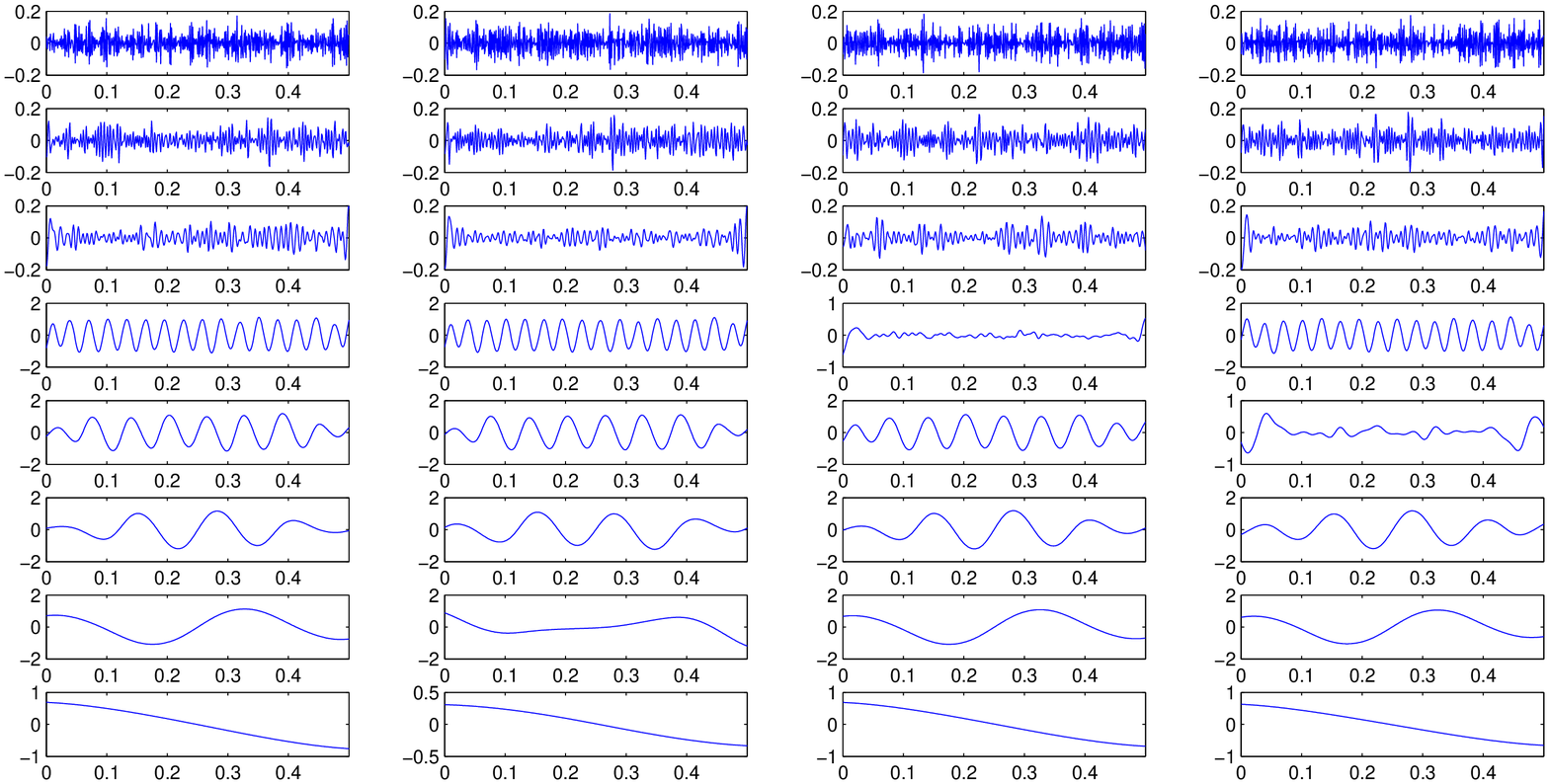}
\caption{ROIMFs obtained from MEMD}
\label{fig:memdROImfs}
\end{figure}

%Matlab code of MEMD used in simulations has been taken from \url{http://www.commsp.ee.ic.ac.uk/~mandic/research/emd.htm}.
We used 4-variate time series signal, which is summation of sinusoids (with combination of frequencies $f_1 =4 Hz, f_2 =8 Hz, f_3 =16 Hz, f_4 =32 Hz$) and the Gaussian white noise of mean 0 and standard deviation of 0.1., i.e.
 \begin{equation}
 x_j(t)=\sum_{i=1}^{4} sin(2\pi f_{i}t)+n_j(t) \qquad \text{ for } j=[1,4]), \label{exam10}
 \end{equation}
for the simulation results shown in Figure~\ref{fig:memdImfs} to \ref{fig:memdROImfs}. The ROIMFs preserve properties of IMF and energy of signal in decomposition.
\begin{table}[!t]
\caption{$Pee$ for IMFs, FOIMFs and ROIMFs obtained from MEMD}
\centering % used for centering table
\begin{tabular}{l l l l} % centered columns (4 columns)
\hline %inserts double horizontal lines
& IMFs & FOIMFs & ROIMFs \\  % inserts table
\hline % inserts single horizontal line
$Pee$ $x_1(t)$ & 23.7398 & 1.1505e-14 & 5.7526e-14\\
$Pee$ $x_2(t)$ & 20.2972 & -2.9996e-14 & 2.9996e-14\\
$Pee$ $x_3(t)$ & 29.9836 & -7.5779e-14 & -4.5467e-14\\
$Pee$ $x_4(t)$ & 19.0329 & -4.5440e-14 & -1.5147e-14\\
\hline %inserts single line
\end{tabular}
\label{table:memdEr} % is used to refer this table in the text
\end{table}
\subsection{The Time-Frequency Analysis of chirp signal}
\begin{figure}[!t]
\centering
\includegraphics[angle=0,width=0.99\textwidth]{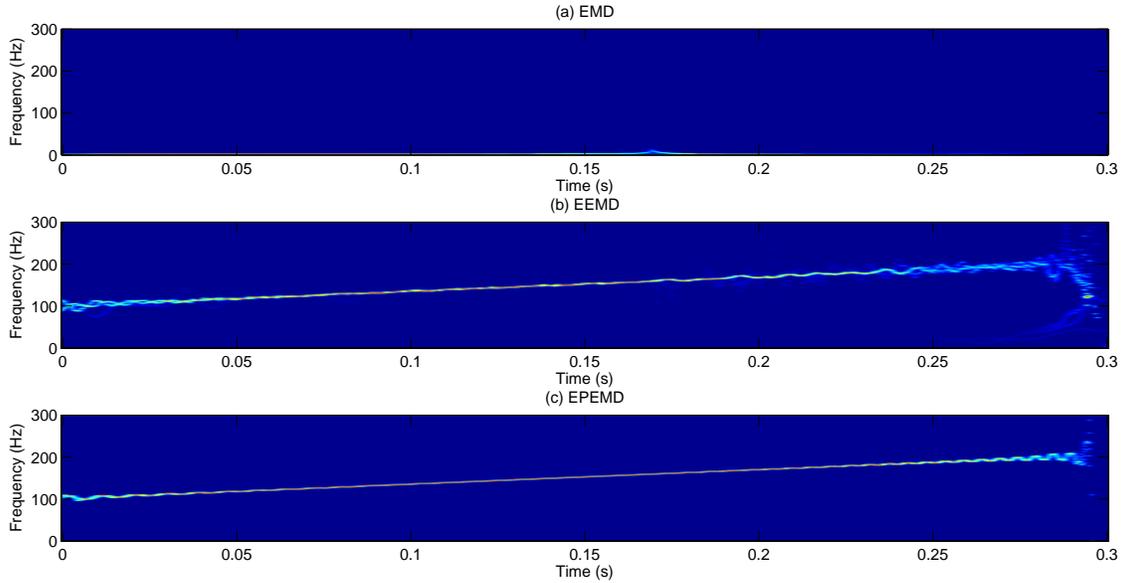}
\caption{Time-Frequency Analysis of linear chirp with zero padding.}
\label{fig:TF_chirp}
\end{figure}

Figure \ref{fig:TF_chirp} shows the Time-Frequency (T-F) estimates for linear chirp (time 0 to 0.3 second, $Fs=10000$ Hz, frequency range 100 to 200 Hz, 50 zero padded), obtained using the EMD, EEMD and EPEMD. There is enhanced T-F tracking when using EPEMD as compare to EMD and EEMD. The reason for the artifacts in EMD and EEMD is high energies in the IMFs of low frequencies due to leakage. The percentage energy leakage (Pee) for EMD, EEMD and EPEMD are $-1.014\times 10^7$, $113.54$ and $3.181\times 10^{-13}$, respectively.
\subsection{The statistical significance of IMFs generated by the proposed EPEMD algorithms}
The statistical significance test of IMFs is developed in \cite{rs10} to determine if data or its IMF components contain relevant and useful or not so relevant information. The IMF components with their energy located within the upper and lower bounds are considered as components generated from the Gaussian white noise present in data and contain no signal information, and the IMF components with their energy located outside the upper and lower bounds contain signal information, at the selected confidence level.

\begin{figure}[!t]
\centering
\includegraphics[angle=0,width=0.99\textwidth]{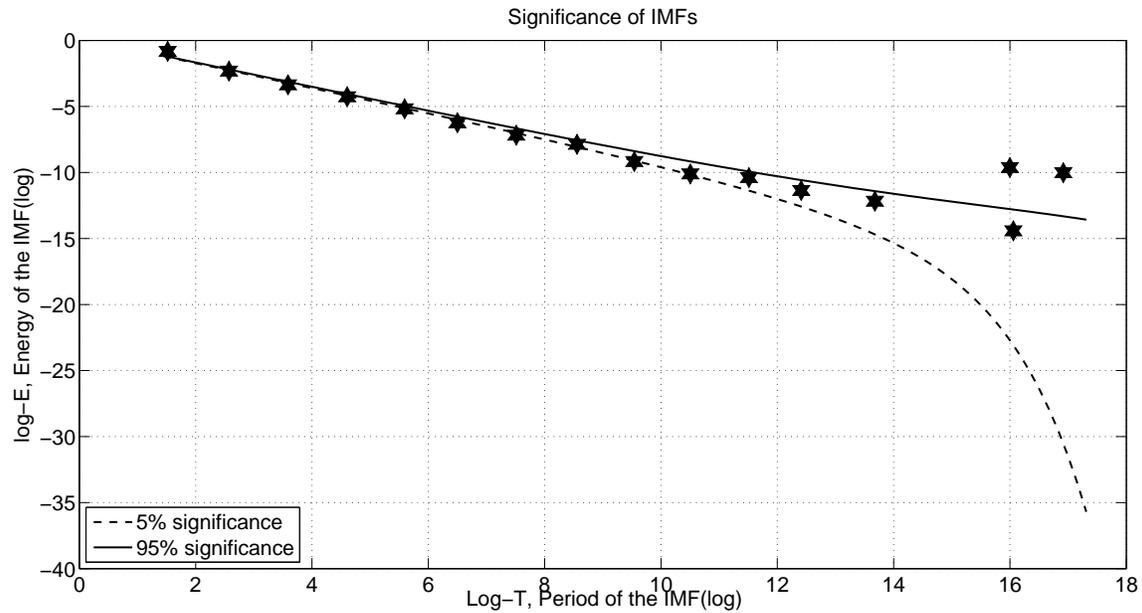}
\caption{Significance test of the IMFs of the Gaussian white noise.}
\label{fig:IMFs_wn_st}
\end{figure}

\begin{figure}[!t]
\centering
\includegraphics[angle=0,width=0.99\textwidth]{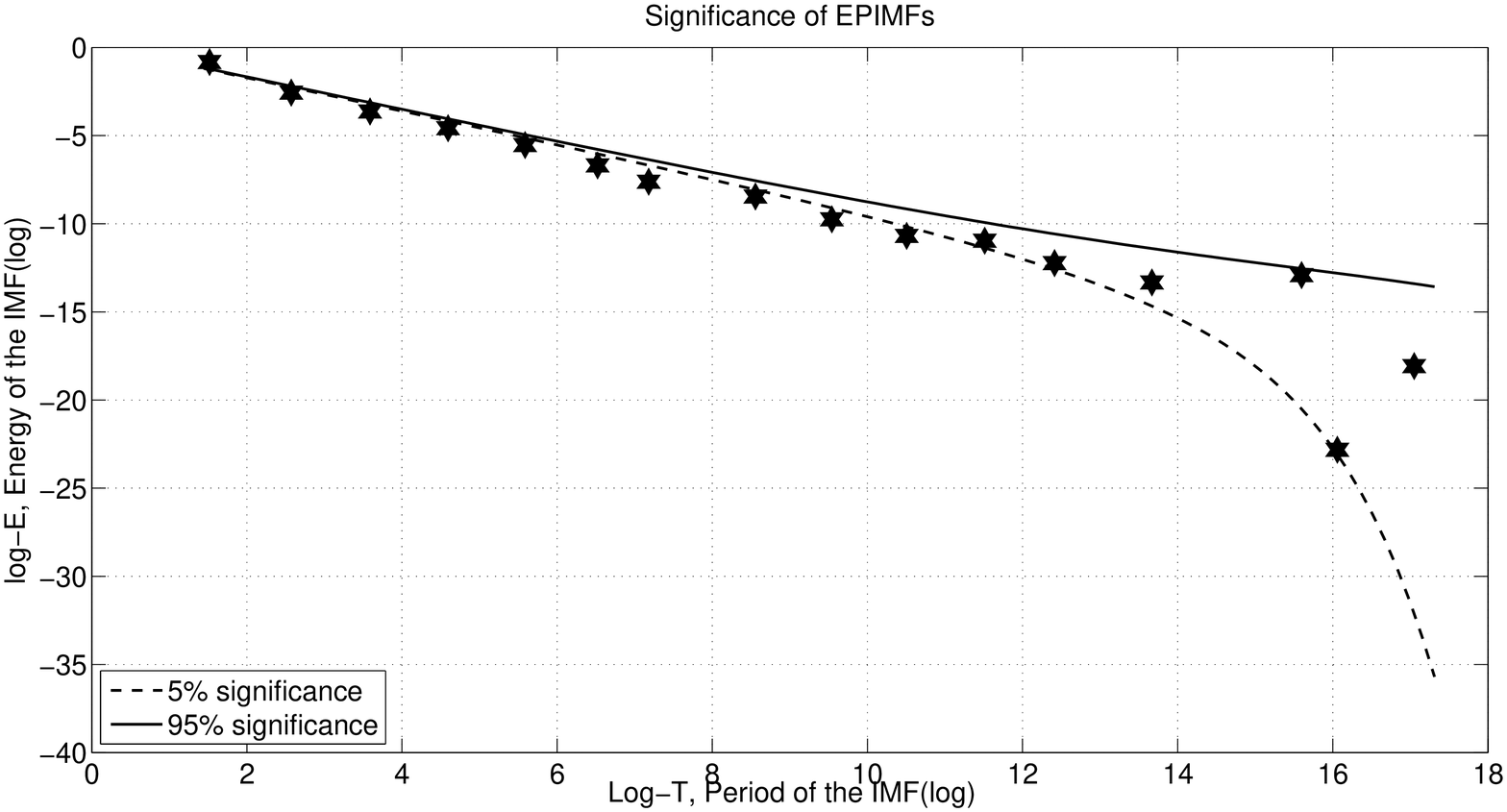}
\caption{Significance test of the EPIMFs of the Gaussian white noise.}
\label{fig:EPIMFs_wn_st}
\end{figure}

\begin{figure}[!t]
\centering
\includegraphics[angle=0,width=0.99\textwidth]{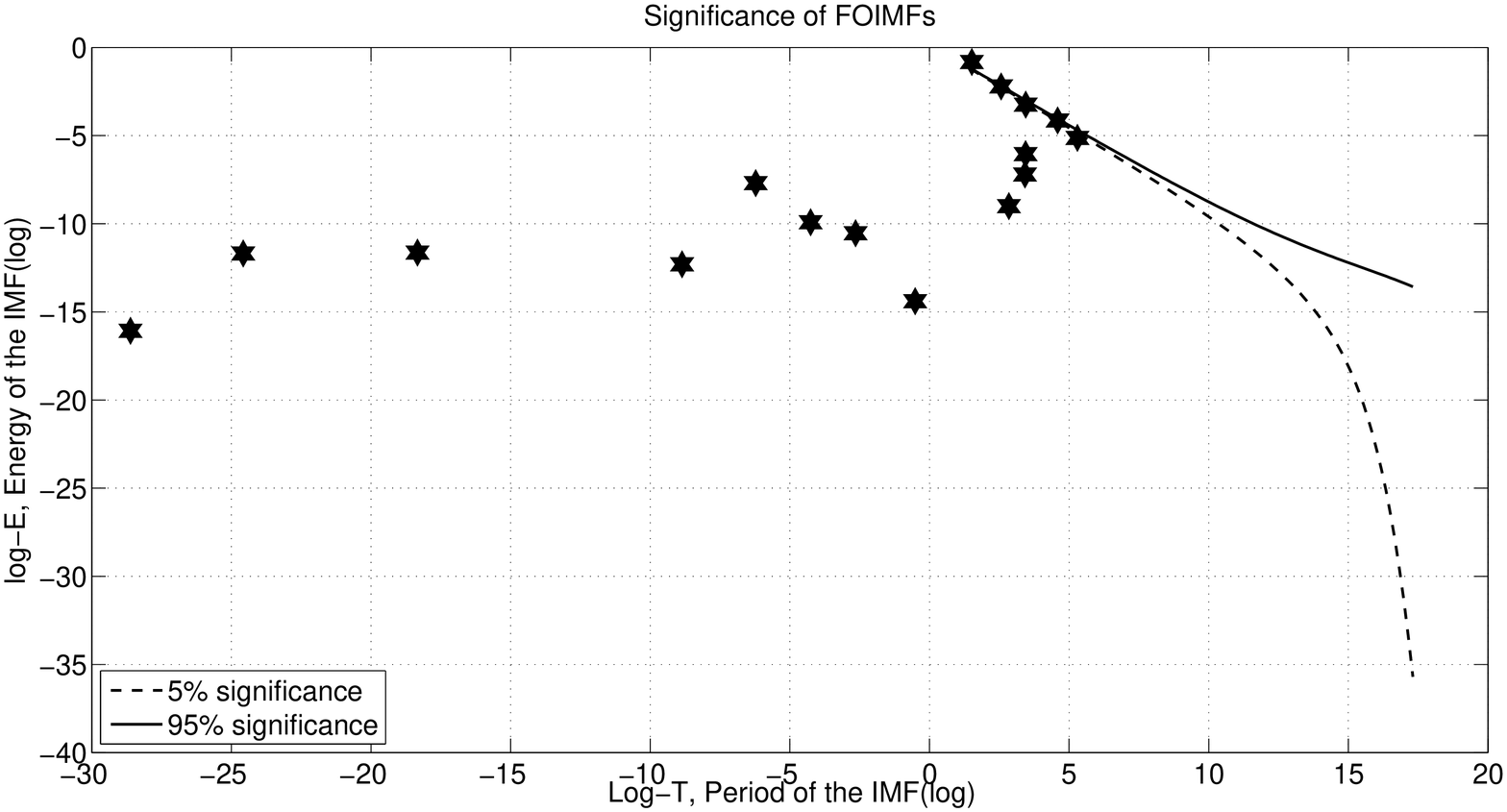}
\caption{Significance test of the FOIMFs of the Gaussian white noise.}
\label{fig:FOIMFs_wn_st}
\end{figure}

\begin{figure}[!t]
\centering
\includegraphics[angle=0,width=0.99\textwidth]{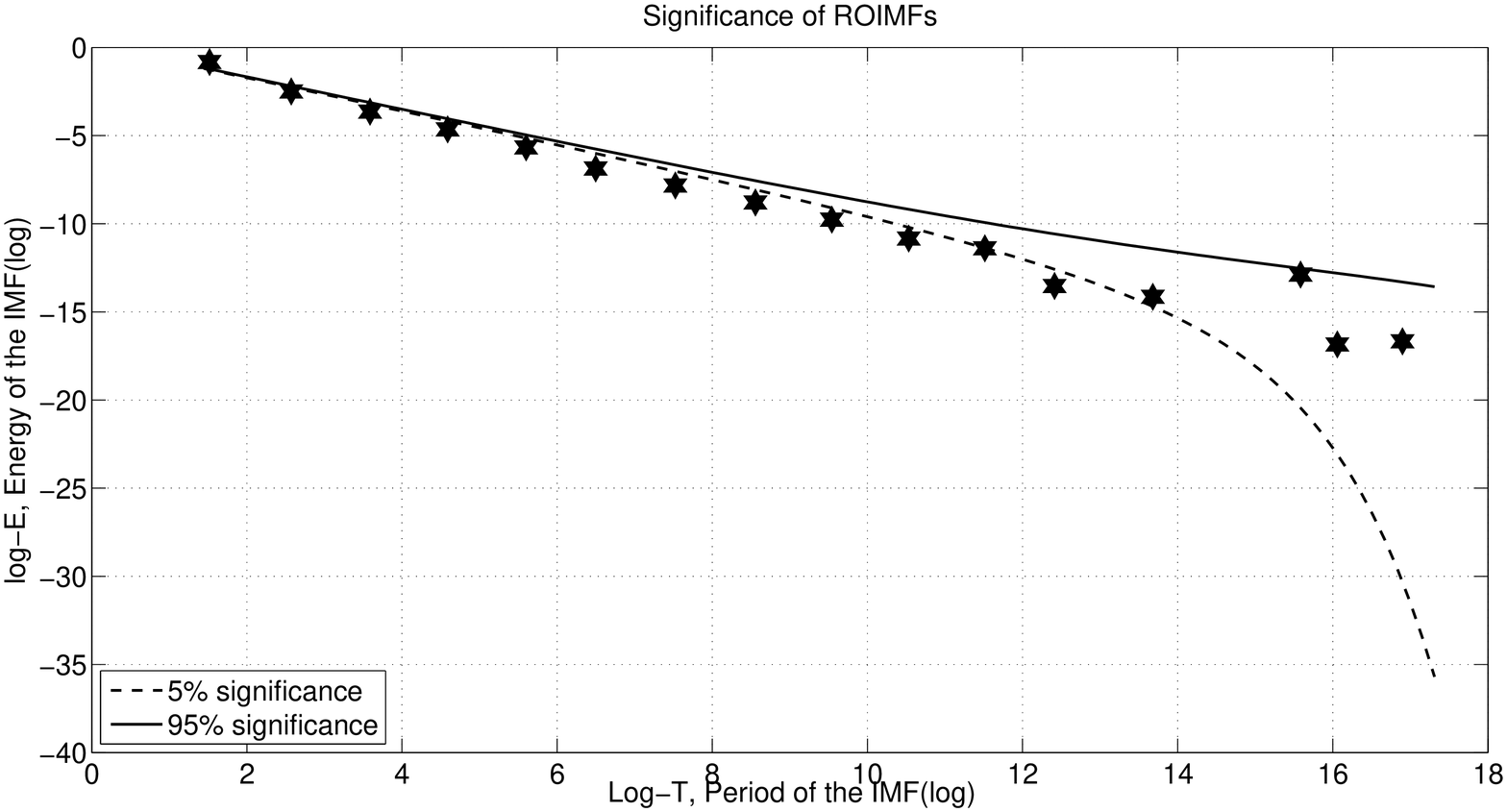}
\caption{Significance test of the ROIMFs of the Gaussian white noise.}
\label{fig:ROIMFs_wn_st}
\end{figure}

\begin{figure}[!t]
\centering
\includegraphics[angle=0,width=0.99\textwidth]{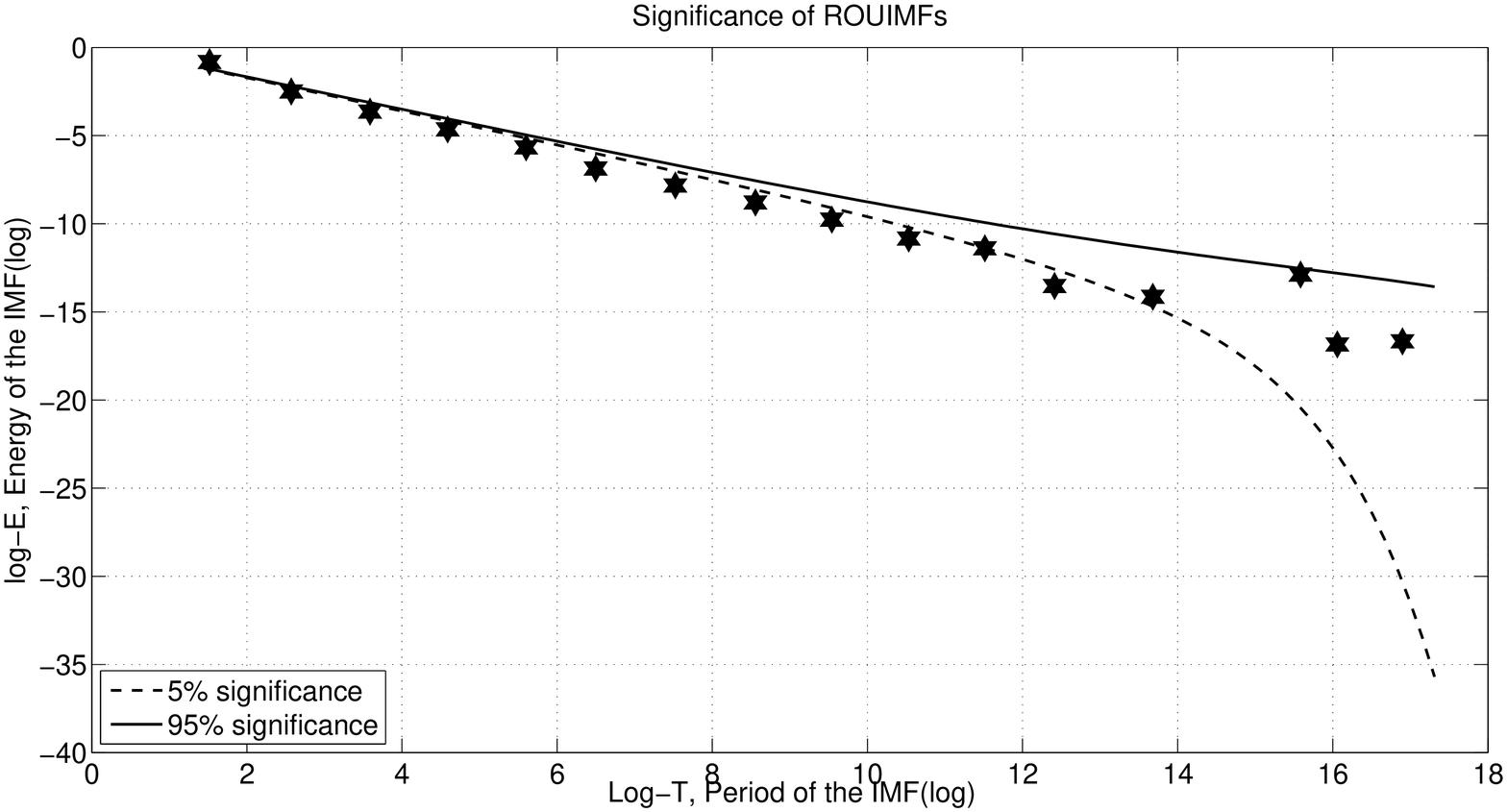}
\caption{Significance test of the ROUIMFs of the Gaussian white noise.}
\label{fig:ROUIMFs_wn_st}
\end{figure}

To study the statistical characteristics of white noise using EMD, the Fourier spectra and mean periods estimation, the statistical significance test for each IMFs has been performed in \cite{rs10}. Similarly, here we obtain Figures \ref{fig:IMFs_wn_st} to \ref{fig:ROUIMFs_wn_st} which are the spread function plot, obtained via EMD and EPEMD, for statistical significance test of the IMFs, EPIMFs, FOIMFs and ROUIMFs of the Gaussian white noise of $2^{16}$ samples with the normal distribution of mean 0 and standard deviation 1. The dashed and solid lines are the $5^{th}$ and $95^{th}$  percentiles, respectively. The stars correspond to the pairs of the averaged mean energy density and the averaged mean period of IMFs. Since we have obtained these plots from the decomposition of the Gaussian white noise, we expect averaged mean energy density well within confidence limit. It is clear from the figures that the averaged mean energy density of all EPIMFs, ROIMFs and ROUIMFs are well within confidence limit and represent the Gaussian white noise, whereas the averaged mean energy density of two IMFs and many FOIMFs components are not within confidence limit and hence these components represent spurious signal components and they are not providing any physical meaning.
\section{Conclusions}
In this paper, we have proposed two energy preserving EMD (EPEMD) algorithms. The first EPEMD algorithm, to preserve the energy of a signal in decomposition, decomposes a signal into the linearly independent (LI), non orthogonal yet energy preserving (LINOEP) IMFs and residue (EPIMFs). In the second algorithm, we have shown that if orthogonalization process proceeds from the lowest frequency component to highest frequency IMF, the GSOM yields functions which preserve the properties of IMFs as well as the energy in decomposition, and hence eliminate the energy leakage among IMFs. The suitability of the generated reverse orthogonal IMFs (ROIMFs) and reverse orthogonal and uncorrelated IMFs (ROUIMFs) are validated through the decomposition of various simulated as well as real life time series. The overall and partial index of orthogonality and energy leakage are used to demonstrate the marked improvement in the orthogonality of the ROIMFs and ROUIMFs components. Finally, the statistical significance test of IMFs, EPIMFs, ROIMFs and ROUIMFs generated from the white Gaussian noise via EMD and EPEMD algorithms are used to illustrate the relevance, improvement and effectiveness of the proposed methodologies. The EPIMFs, ROIMFs and ROUIMFs generated by various EMDs preserve the properties of IMFs and the energy of signal in the decomposition, and are promising and generating better results.
\section*{Acknowledgment}
The authors would like to thank JIIT Noida, for permitting to carry out research at IIT, Delhi and providing all required resources throughout this study.
%(IITD). The authors also would like to thank IITD for providing resources and technical supports required throughout this
%study.
\ifCLASSOPTIONcaptionsoff
  \newpage
\fi

\end{document}